\documentclass[twoside,twocolumn,9pt]{article}
\usepackage{extsizes}
\usepackage[super,sort&compress,comma]{natbib} 
\usepackage[version=3]{mhchem}
\usepackage[left=1.5cm, right=1.5cm, top=1.785cm, bottom=2.0cm]{geometry}
\usepackage{balance}
\usepackage{times,mathptmx}
\usepackage{sectsty}
\usepackage{graphicx} 
\usepackage{lastpage}
\usepackage[format=plain,justification=justified,singlelinecheck=false,font={stretch=1.125,small,sf},labelfont=bf,labelsep=space]{caption}
\usepackage{float}
\usepackage{fancyhdr}
\usepackage{fnpos}
\usepackage[english]{babel}
\usepackage{array}
\usepackage{droidsans}
\usepackage{charter}
\usepackage[T1]{fontenc}
\usepackage[usenames,dvipsnames]{xcolor}
\usepackage{setspace}
\usepackage{multirow}
\usepackage[compact]{titlesec}

\usepackage{epstopdf}

\definecolor{cream}{RGB}{222,217,201}

\begin{document}

\pagestyle{fancy}
\thispagestyle{plain}
\fancypagestyle{plain}{

\fancyhead[C]{\includegraphics[width=18.5cm]{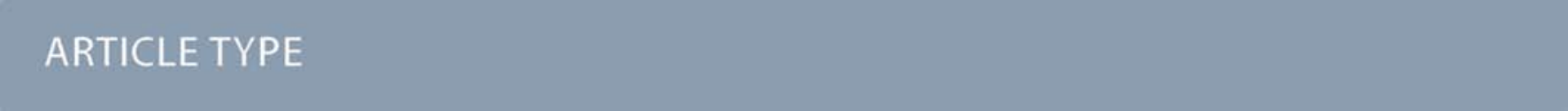}}
\fancyhead[L]{\hspace{0cm}\vspace{1.5cm}\includegraphics[height=30pt]{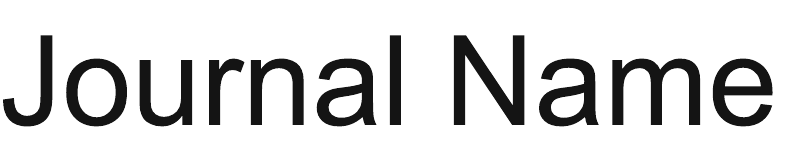}}
\fancyhead[R]{\hspace{0cm}\vspace{1.7cm}\includegraphics[height=55pt]{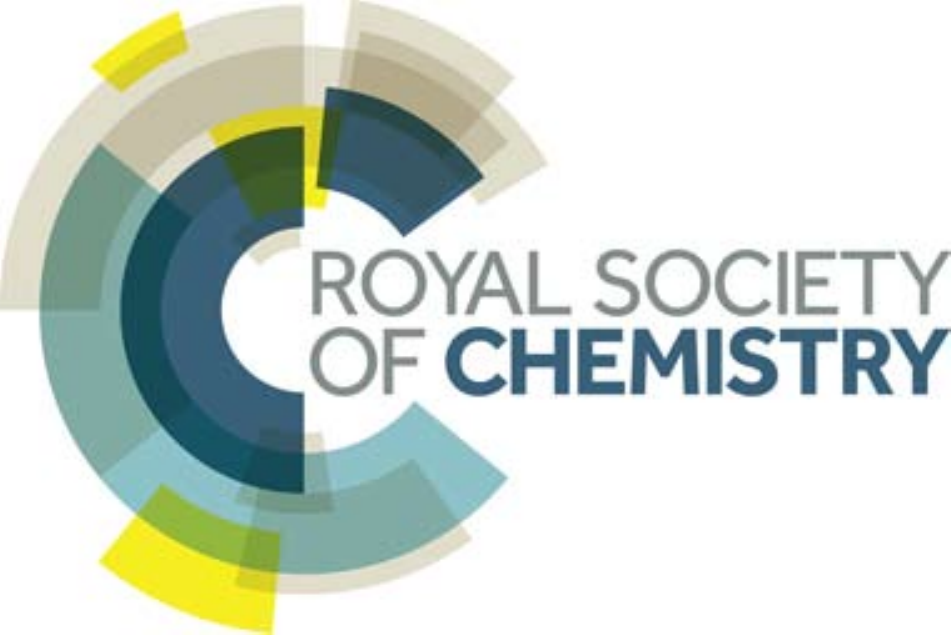}}
\renewcommand{\headrulewidth}{0pt}
}

\makeFNbottom
\makeatletter
\renewcommand\LARGE{\@setfontsize\LARGE{15pt}{17}}
\renewcommand\Large{\@setfontsize\Large{12pt}{14}}
\renewcommand\large{\@setfontsize\large{10pt}{12}}
\renewcommand\footnotesize{\@setfontsize\footnotesize{7pt}{10}}
\makeatother

\renewcommand\floatpagefraction{.99}
\renewcommand\topfraction{.99}
\renewcommand\bottomfraction{.99}
\renewcommand\textfraction{.01}
\renewcommand\dbltopfraction{0.99}
\renewcommand\dblfloatpagefraction{0.99}

\renewcommand{\thefootnote}{\fnsymbol{footnote}}
\renewcommand\footnoterule{\vspace*{1pt}%
\color{cream}\hrule width 3.5in height 0.4pt \color{black}\vspace*{5pt}} 
\setcounter{secnumdepth}{5}

\makeatletter 
\renewcommand\@biblabel[1]{#1}            
\renewcommand\@makefntext[1]%
{\noindent\makebox[0pt][r]{\@thefnmark\,}#1}
\makeatother 
\renewcommand{\figurename}{\small{Fig.}~}
\sectionfont{\sffamily\Large}
\subsectionfont{\normalsize}
\subsubsectionfont{\bf}
\setstretch{1.125} 
\setlength{\skip\footins}{0.8cm}
\setlength{\footnotesep}{0.25cm}
\setlength{\jot}{10pt}
\titlespacing*{\section}{0pt}{4pt}{4pt}
\titlespacing*{\subsection}{0pt}{15pt}{1pt}

\fancyfoot{}
\fancyfoot[LO,RE]{\vspace{-7.1pt}\includegraphics[height=9pt]{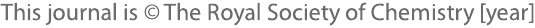}}
\fancyfoot[CO]{\vspace{-7.1pt}\hspace{13.2cm}\includegraphics{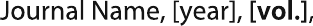}}
\fancyfoot[CE]{\vspace{-7.2pt}\hspace{-14.2cm}\includegraphics{head_foot/RF}}
\fancyfoot[RO]{\footnotesize{\sffamily{1--\pageref{LastPage} ~\textbar  \hspace{2pt}\thepage}}}
\fancyfoot[LE]{\footnotesize{\sffamily{\thepage~\textbar\hspace{3.45cm} 1--\pageref{LastPage}}}}
\fancyhead{}
\renewcommand{\headrulewidth}{0pt} 
\renewcommand{\footrulewidth}{0pt}
\setlength{\arrayrulewidth}{1pt}
\setlength{\columnsep}{6.5mm}
\setlength\bibsep{1pt}

\makeatletter 
\newlength{\figrulesep} 
\setlength{\figrulesep}{0.5\textfloatsep} 

\newcommand{\topfigrule}{\vspace*{-1pt}%
\noindent{\color{cream}\rule[-\figrulesep]{\columnwidth}{1.5pt}} }

\newcommand{\botfigrule}{\vspace*{-2pt}%
\noindent{\color{cream}\rule[\figrulesep]{\columnwidth}{1.5pt}} }

\newcommand{\dblfigrule}{\vspace*{-1pt}%
\noindent{\color{cream}\rule[-\figrulesep]{\textwidth}{1.5pt}} }

\makeatother

\twocolumn[
  \begin{@twocolumnfalse}
\vspace{3cm}
\sffamily
\begin{tabular}{m{4.5cm} p{13.5cm} }

\includegraphics{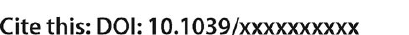} & \noindent\LARGE{\textbf{Negative thermal expansion and metallophilicity in Cu$_{\textrm3}$[Co(CN)$_{\textrm6}$]$^\dag$}} \\
\vspace{0.3cm} & \vspace{0.3cm} \\
 & \noindent\large{Adam F. Sapnik,\textit{$^{a}$} Xiaofei Liu,\textit{$^{b}$} Hanna L. B. Bostr{\"o}m,\textit{$^{a}$} Chloe S. Coates,\textit{$^{a}$} Alistair R. Overy,\textit{$^{a,c}$} Emily M. Reynolds,\textit{$^{a}$} Alexandre Tkatchenko,\textit{$^{d}$} and Andrew L. Goodwin$^{\ast}$\textit{$^{a}$}} \\
\includegraphics{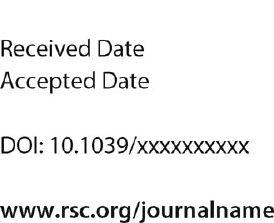} & \noindent\normalsize{We report the synthesis and structural characterisation of the molecular framework copper(I) hexacyanocobaltate(III), Cu$_{\textrm3}$[Co(CN)$_{\textrm6}$], which we find to be isostructural to H$_{\textrm3}$[Co(CN)$_{\textrm6}$] and the colossal negative thermal expansion material Ag$_{\textrm3}$[Co(CN)$_{\textrm6}$]. Using synchrotron X-ray powder diffraction measurements, we find strong positive and negative thermal expansion behaviour respectively perpendicular and parallel to the trigonal crystal axis: $\alpha_a$ = 25.4(5)\,MK$^{-1}$ and $\alpha_c$ = $-$43.5(8)\,MK$^{-1}$. These opposing effects collectively result in a volume expansivity $\alpha_V$ = 7.4(11)\,MK$^{-1}$ that is remarkably small for an anisotropic molecular framework. This thermal response is discussed in the context of the behaviour of the analogous H- and Ag-containing systems. We make use of density-functional theory with many-body dispersion
interactions (DFT+MBD) to demonstrate that Cu$\ldots$Cu metallophilic (`cuprophilic') interactions are significantly weaker in Cu$_3$[Co(CN)$_6$] than Ag$\ldots$Ag interactions in Ag$_{\textrm3}$[Co(CN)$_{\textrm6}$], but that this lowering of energy scale counterintuitively translates to a more moderate---rather than enhanced---degree of structural flexibility. The same conclusion is drawn from consideration of a simple lattice dynamical model, which we also present here. Our results demonstrate that strong interactions can actually be exploited in the design of ultra-responsive materials if those interactions are set up to act in tension.} \\

\end{tabular}

 \end{@twocolumnfalse} \vspace{0.6cm}
 
  ]

\renewcommand*\rmdefault{bch}\normalfont\upshape
\rmfamily
\section*{}
\vspace{-1cm}


\footnotetext{\textit{$^{a}$~Department of Chemistry, University of Oxford, Inorganic Chemistry Laboratory, South Parks Road, Oxford OX1 3QR, U.K. Tel: +44 (0)1865 272137; E-mail: andrew.goodwin@chem.ox.ac.uk}\\
\textit{$^{b}$~State Key Laboratory of Mechanics and Control of Mechanical Structures, Key Laboratory for Intelligent Nano Materials and Devices of Ministry of Education, Nanjing University of Aeronautics and Astronautics, Nanjing 210016, China}\\
\textit{$^{c}$~Diamond Light Source, Chilton, Oxfordshire OX11 0DE, U.K.}\\
\textit{$^{d}$~Physics and Materials Science Research Unit, University of Luxembourg, L-1511 Luxembourg}}

\footnotetext{\dag~Electronic Supplementary Information (ESI) available: X-ray powder diffraction measurements and Rietveld fits, lattice parameter data and structural parameter data, \emph{ab initio} van der Waals parameters and energies, GULP input files. See DOI: 10.1039/b000000x/}


\section{Introduction}

The development of responsive materials often exploits weak interactions as key design elements because lower interaction energies heighten the sensitivity of a material to external perturbations.\cite{Coudert_2015,OCM2012,Horike_2009,Krishna_2016} It is no accident, for example, that the weak inter-molecular forces in molecular crystals generally allow more extreme responses to changes in temperature\cite{Das_2010,Goodwin_2010} and pressure\cite{Shepherd_2012,Duyker_2016} than is possible in conventional inorganic ceramics, the structures of which are held together by strong ionic and covalent bonding networks. In this context, supramolecular interactions assume a particular importance, given that their energy scales are so much lower than those of electrostatic or covalent interactions. Hence the prevalence of hydrogen-bonding,\cite{LTB2014} halogen-bonding,\cite{JKM2014} $\pi$--$\pi$,\cite{Yot_2012} van der Waals (vdW),\cite{Das_2010} host--guest,\cite{ESB2014,Salles_2010} and metallophilic\cite{GKT2008a} interactions amongst many of the important materials in the field.

\begin{figure}[t]
\centering
  \includegraphics[width=8.3cm]{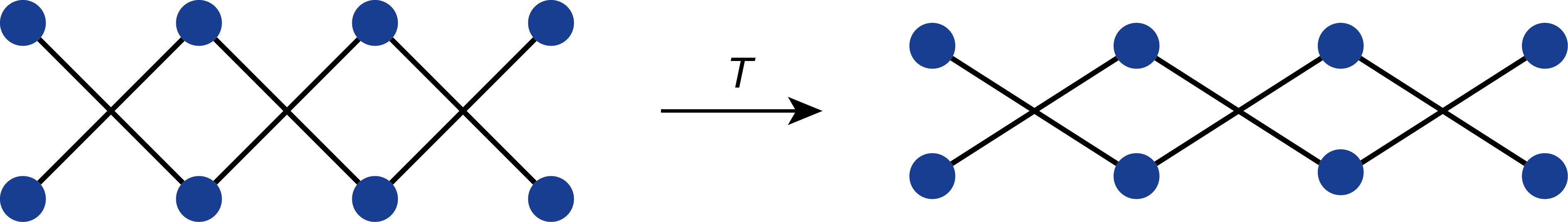}
  \caption{``Wine-rack'' mechanism for anisotropic thermal expansion in flexible framework materials. Horizontal expansion couples to vertical contraction \emph{via} lattice flexing.}
  \label{fig1}
\end{figure}

Thermal expansion behaviour is a straightforward measure of responsiveness: it quantifies the effect of temperature on the linear dimensions of a material.\cite{HT1998} Compounds with large thermal expansion coefficients often show extreme and counterintuitive responses to pressure,\cite{Krishnan_1979,GKT2008b} and may harbour various other anomalous elastic properties, such as negative Poisson's ratios\cite{Greaves_2011} or thermosalient\cite{PRC2014,PCC2016} effects. So it is perhaps unsurprising that some of the most extreme (`colossal') thermal expansion known has been observed in framework materials whose lattice dimensions are a function of weak metallophilic interactions.\cite{GKT2008a,GKT2008b,Korcok_2009} The canonical system of this type is Ag$_3$[Co(CN)$_6$], which adopts a lattice structure\cite{LG1968} that can flex in such a way as to vary argentophilic Ag$\ldots$Ag separations without affecting covalent interactions within the lattice itself.\cite{GCC2008,CGT2008} A geometric consequence of this flexing behaviour is that the positive thermal expansion (PTE) of argentophilic interactions (\emph{i.e.}\ increase in separation with increasing temperature) is translated into a negative thermal expansion (NTE) effect in a perpendicular direction [Fig.~\ref{fig1}]. The same mechanism operates under application of hydrostatic pressure, such that volume compression actually results in linear expansion for a particular set of directions\cite{GKT2008b}---so-called negative linear compressibility (NLC).\cite{BRS1998,Lakes_2008,Cairns_2015} NTE and NLC are valuable material properties, exploitable in the design of athermal composites used in optical devices and next-generation pressure sensors.

In seeking to design even more responsive analogues of Ag$_3$[Co(CN)$_6$], we considered the possibility of replacing Ag by Cu. Metallophilic interactions involving Cu$^+$ ions are perhaps less well studied than argentophilic and aurophilic interactions, but are expected to be weaker given the reduced polarisability of the $3d$ shell.\cite{Jansen_1987,OK2004} Hence, by the arguments discussed above, Cu$_3$[Co(CN)$_6$] has always been an obvious candidate for extreme thermomechanical response. To the best of our knowledge, this system has never previously been reported: the difficulty of preparing the phase is likely a consequence of the propensity for Cu$^+$ to disproportionate under the aqueous reaction conditions used to prepare the family of materials related to Ag$_3$[Co(CN)$_6$].\cite{GCC2008} We have recently exploited the Cu$^{2+}$ reduction protocol developed in Ref.~\citenum{BCD2001} to allow access to otherwise unrealisable Cu(I)-containing frameworks,\cite{Hunt_2015} suggesting that a similar synthetic approach may provide an alternative synthetic entry point to Cu$_3$[Co(CN)$_6$].

Here we validate such an approach, reporting the synthesis, crystal structure, and thermal expansion behaviour of Cu$_3$[Co(CN)$_6$]. Using a combination of high-resolution synchrotron X-ray diffraction measurements and Rietveld refinement, we show the system to be isostructural with Ag$_3$[Co(CN)$_6$] and H$_3$[Co(CN)$_6$].\cite{LG1968,Pauling_1968,Haser_1977,KDE2010} Variable-temperature (100--598\,K) diffraction measurements allow determination of the corresponding coefficients of thermal expansion $\alpha_\ell=(\partial\ln\ell/\partial T)_p$, which we find to be substantially \emph{less} extreme than those of Ag$_3$[Co(CN)$_6$] (even if they remain large in the context of the behaviour conventional inorganic solids\cite{Lind_2012}). In particular, our data give $\alpha_a=25.4(5)$\,MK$^{-1}$ and $\alpha_c = -43.5(8)$\,MK$^{-1}$; \emph{cf} $\alpha_a=144(9)$\,MK$^{-1}$ and $\alpha_c = -126(4)$\,MK$^{-1}$ for Ag$_3$[Co(CN)$_6$].\cite{GCC2008} In order to rationalise this more moderate thermomechanical response in terms of the relative strengths of Cu$^+\ldots$Cu$^+$ and Ag$^+\ldots$Ag$^+$ metallophilic interactions, we carry out a series of \emph{ab initio} calculations. Our analysis suggests (i) that cuprophilic interactions are indeed weaker than argentophilic interactions in this family, and (ii) the more extreme thermomechanical response of the Ag-containing compound is a result of the balance of metallophilic and electrostatic interaction energies rather than a signature of particularly weak argentophilicity. Lattice dynamical calculations using a highly simplified interaction model relevant to the entire A$_3$[Co(CN)$_6$] structural family lead to the same conclusions. Our results suggest that \emph{competing} interactions---rather than low-energy interactions \emph{per se}---might be key in the design of ultra-responsive materials.

\section{Methods}

All reagents were obtained from commercial suppliers and used as received.

\subsection{Copper(I) hexacyanocobaltate(III)}

We prepared polycrystalline samples of copper(I) hexacyanocobaltate(III) following a modification of the reduction protocol reported in Refs.~\citenum{BCD2001,Hunt_2015}. A saturated aqueous solution of copper(II) sulfate (Sigma Aldrich, 99\%; 0.17705 g) was added dropwise to a concentrated aqueous solution of sodium bisulfite (Sigma Aldrich, 0.05771 g), present in stoichiometric excess, to afford a mint-green solution. The solution was stirred for 30\,min, after which time an aqueous solution of potassium hexacyanocobaltate(III) (Sigma Aldrich, 97\%, 0.12288 g; stoichiometric with respect to copper) was added dropwise to afford a pale blue precipitate. The solution was stirred for a further 2\,h, and the pale-blue solid product formed was isolated by filtration, washing (H$_2$O) and drying under vacuum. The solid contained a mixture of copper(I) hexacyanocobaltate(III) and Prussian-blue-structured potassium copper(II) hexacyanocobaltate(III), a seemingly inescapable by-product of this synthetic strategy.

Copper(I) hexacyanocobaltate(III) could also be obtained in impure form using mechanochemical synthesis. Stoichiometric quantities of solid tetrakis(acetonitrilo)copper(I) hexafluorophosphate (Chem Cruz, 98\%, 0.41346) and potassium hexacyanocobaltate (Sigma Aldrich, 97\%, 0.12288g) were combined in an agate mortar, and intimately mixed \emph{via} solid-state grinding for 30\,min. An obvious colour change from white to pale blue occurred during this process. The resulting solid was washed (H$_2$O) and dried to afford a mixture of copper(I) hexacyanocobaltate(III),  potassium copper(II) hexacyanocobaltate(III) and at least one further unidentified product. Given the reduced purity of this product, the solution-phase product described above was used for all diffraction measurements carried out in this study. 

\subsection{Powder X-ray diffraction}

High-resolution synchrotron powder diffraction measurements were carried out using the I11 beamline at the Diamond Light Source. A finely-ground sample of copper(I) hexacyanocobaltate(III), prepared as above, was loaded into a borosilicate capillary (0.5\,mm diameter) and mounted on the diffractometer. Diffraction patterns were recorded using the Mythen2 point sensitive detector over the angular range $2\theta=2$--$92^\circ$, using an X-ray wavelength $\lambda=0.826210$\,\AA\ calibrated by refinement of a silicon NIST 640c standard. Each measurement consisted of two scans of 5\,s exposure, offset relative to one another by $\Delta2\theta=0.25^\circ$. The sample temperature was controlled using an Oxford Cryostream (100--500\,K) and a Cyberstar hot air blower (523--598\,K). Diffraction patterns were measured at intervals of 25\,K between 100 and 500\,K and again between 523 and 598\,K.

Both Pawley and Rietveld refinements were carried out using TOPAS Academic (version 4.1).\cite{topas} We employed a modified Thompson--Cox--Hasting pseudo-Voigt (TCHZ) peak shape, combined with a simple axial divergence correction and a Stephens anisotropic peak broadening term.\cite{SP1999} The potassium copper(II) hexacyancobaltate(III) impurity phase was modelled using Pawley refinement of the $Fm\bar3m$ double-metal cyanide cell ($a\sim$10\,\AA).\cite{RM2000} Rietveld refinement of the Cu$_3$[Co(CN)$_6$] phase made use of a starting model based on the known structure of Ag$_3$[Co(CN)$_6$].\cite{LG1968} Refinement was stable for all temperature points, provided that Co--C/C--N bond distance restraints and a single isotropic displacement parameter for all atom types were used in the Rietveld model.  Sequential (seed-batch) Rietveld refinements, where the starting structural parameters for each temperature point were those used at the preceding temperature, provided structural models with physically-sensible temperature dependencies for $T\leq450$\,K. For the temperature regime $450\leq T\leq598$\,K, we found that the positional coordinates of the C and N atoms and the value of $B_{\textrm{iso}}$ showed strong covariance, and hence we have reduced confidence in the absolute values of these parameters. This regime corresponds to the temperature range over which decomposition of the KCu[Co(CN)$_6$] phase appears to set in.

\subsection{Thermal expansivity determination}

Thermal expansivities were calculated using the PASCal software.\cite{pascal} We employed estimated temperature uncertainties of 5\,K and fitted the principal axis expansivities using linear functions. For internal consistency with the uniaxial expansivities, the volume expansivity was determined using the trace of the expansivity tensor\cite{Nye_1957} rather than \emph{via} the direct $V$--$T$ fit given by PASCal.\cite{pascal}

\subsection{\emph{Ab initio} calculations}
\emph{Ab initio} calculations were performed within the FHI-aims code,\cite{Blum_2009} using the numeric atom-centred orbital tier 1 basis set for the wavefunction and a $5\times5\times5$ $k$-point mesh for the Brillouin zone sampling. The Perdew-Burke-Ernzernhof (PBE) functional\cite{Perdew_1996} was used to model the semilocal exchange-correlation energy. To describe the non-local dispersion energies, we used both the interatomic pairwise Tkatchenko-Scheffler (TS) method,\cite{Tkatchenko_2009} as well as the many-body dispersion (MBD) method, which includes many-body dipolar interatomic interactions to all orders in perturbation theory.\cite{Tkatchenko_2012,Ambrosetti_2014} The lattice constants were obtained from unit cell relaxations with cell angles fixed to experimental values. Full \emph{a posteriori} relaxation of the unit cell proved the reliability of this scheme.

\subsection{Lattice dynamical calculations}

Lattice dynamical calculations made use of the GULP software (version 4.4).\cite{gulp} Cell optimisations were carried out under constant pressure conditions $p=0$ and at $T=0$, with strains constrained by symmetry. Dispersion interactions were modelled using a Buckingham potential with vanishingly small repulsive term, and the `{\tt c6}' flag was activated to employ Ewald-type summation. For all calculations, checks were carried out to ensure optimisation convergence and to verify the conservation of angle terms in the parameterisation.

\section{Results and discussion}
\subsection{Crystal structure of Cu$_{\textbf3}$[Co(CN)$_{\textbf6}$]}

\begin{figure}[b]
\centering
  \includegraphics{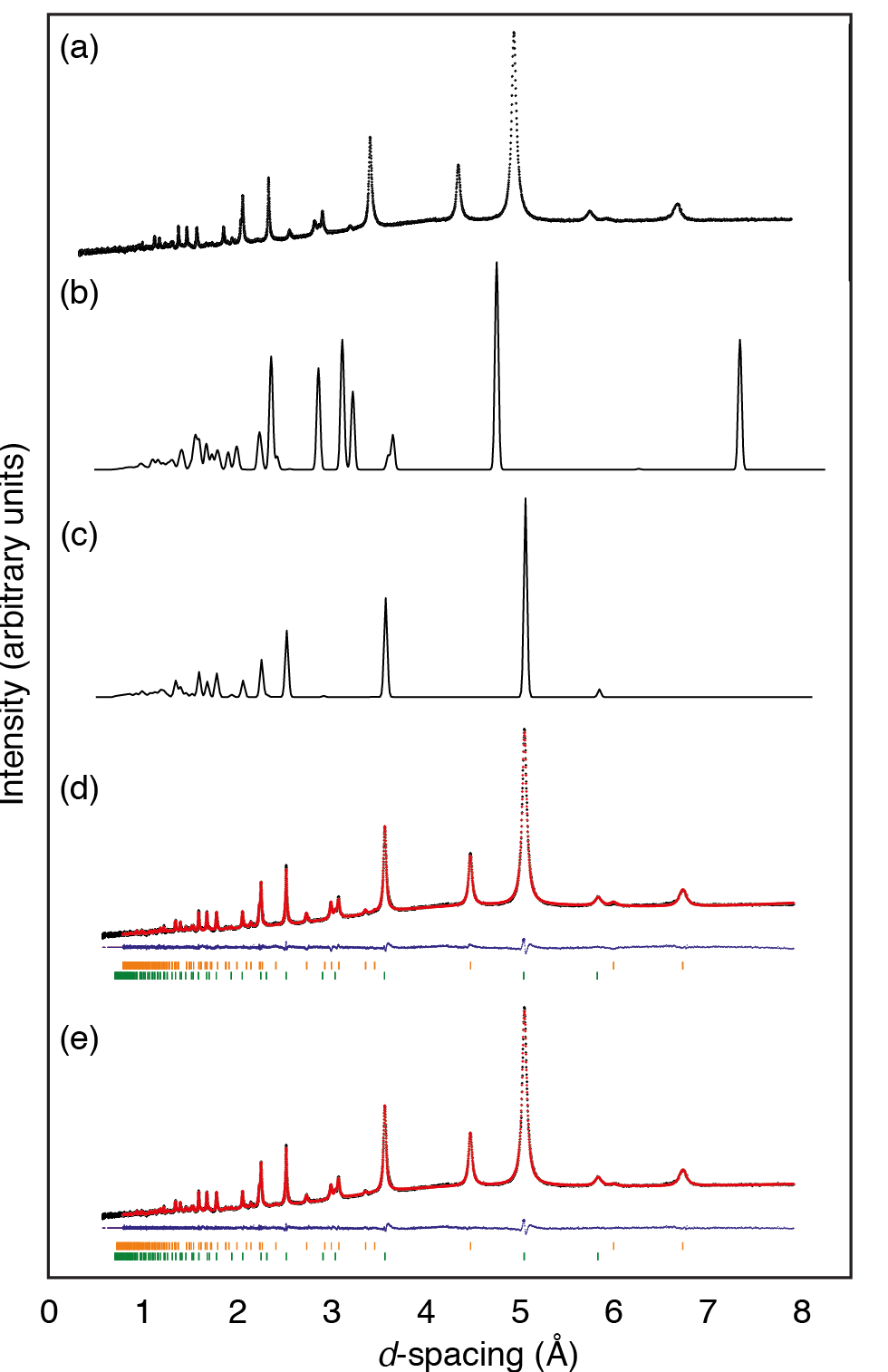}
  \caption{X-ray powder diffraction behaviour and its interpretation in Cu$_3$[Co(CN)$_6$]. (a) Experimental X-ray diffraction pattern, containing reflections attributable to two separate phases: one related to that of the trigonal phase Ag$_3$[Co(CN)$_6$] (diffraction pattern calculated from the model of Ref.~\citenum{LG1968} shown in (b)), and one corresponding to the Prussian blue analogue KCu[Co(CN)$_6$] (Ref.~\citenum{RM2000}), shown in (c). Corresponding two-phase (d) Pawley and (e) Rietveld refinements as described in the text. Data shown as black points, calculated diffraction patterns shown as solid black lines, Pawley/Rietveld fits shown as red points, and difference functions (data $-$ fit) shown as solid blue lines. Tick marks denote the positions of symmetry-allowed reflections for the Cu$_3$[Co(CN)$_6$] (orange) and KCu[Co(CN)$_6$] (green) phases.}
  \label{fig2}
\end{figure}

The ambient-temperature X-ray powder diffraction pattern of our Cu$_3$[Co(CN)$_6$] sample is shown in Fig.~\ref{fig2}, where it is compared to that of Ag$_3$[Co(CN)$_6$] as reported in Ref.~\citenum{GCC2008}. The structural similarity between the two phases is immediately evident, as is the presence of a substantial quantity of an impurity phase. We could account for the entire diffraction pattern using two components, one based on the Ag$_3$[Co(CN)$_6$] structure-type (space group symmetry $P\bar31m$) and one with the cubic Prussian blue structure (space group symmetry $Fm\bar3m$). This second phase would be consistent with the formation of KCu[Co(CN)$_6$] during synthesis, which is certainly feasible on chemical grounds.\cite{Sharpe_1976,WKW2005} A Pawley fit using this two-phase model confirms our assignment of space group symmetries and rules out the presence of any additional crystalline phases [Fig.~\ref{fig2}(d)]. We note that there is good (if fortuitous) distinction between the diffraction profiles of the two phases present, which allows us to clearly distinguish the corresponding lattice parameters and their thermal expansion behaviour (see SI).

\begin{figure}
\centering
  \includegraphics{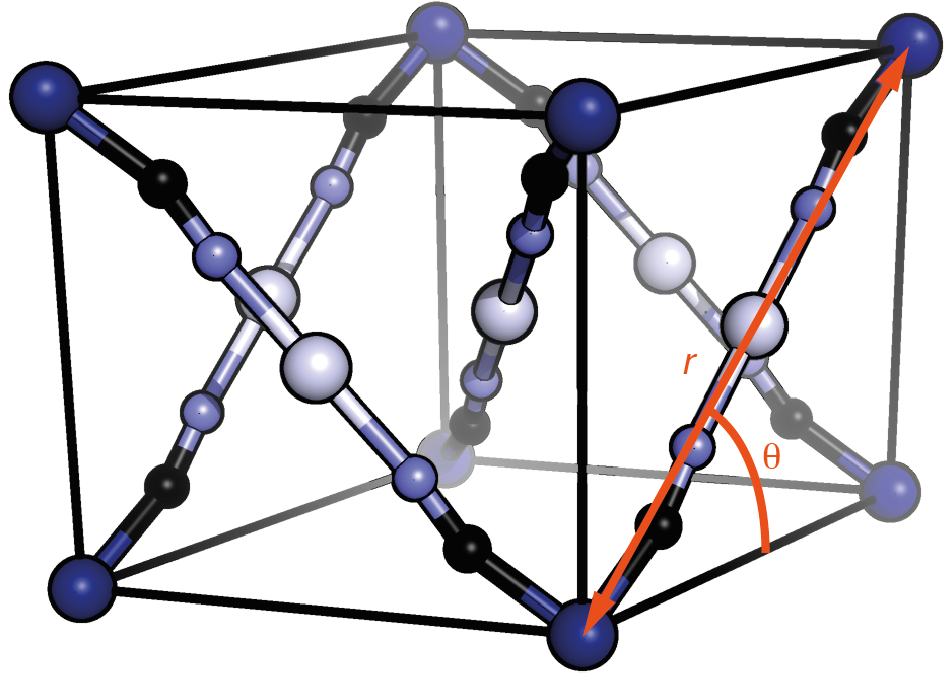}
  \caption{Structural model for Cu$_3$[Co(CN)$_6$] determined using Rietveld refinement of X-ray powder diffraction data collected at 298\,K. Co atoms shown in dark blue, Cu atoms in blue--white, N atoms in blue, and C atoms in black. The XBUs $r$ and $\theta$---shown here in orange---correspond to the framework strut length and hingeing angle, respectively.}
  \label{fig3}
\end{figure}

Having established the space group symmetry of Cu$_3$[Co(CN)$_6$] we proceeded to carry out a Rietveld refinement, employing a starting model based on the lattice parameters obtained during Pawley fitting and the published atom coordinates of Ag$_3$[Co(CN)$_6$].\cite{LG1968} We continued to model the KCu[Co(CN)$_6$] phase using a Pawley fit---indeed this is the case for all subsequent refinements and is not discussed further. We found good stability in the refinement of our structural model of Cu$_3$[Co(CN)$_6$], obtaining a $R$-value of $3.029\%$; the corresponding fit is shown in Fig.~\ref{fig2}(e) and the relevant structural details are summarised in Table~\ref{table1}. A representation of the crystal structure itself is given in Fig.~\ref{fig3}. All refined bond lengths are chemically sensible: we find a Co--C distance of 1.832(11)\,\AA, which is similar to that in Ag$_3$[Co(CN)$_6$] ($d($Co--C$)=1.895$\,\AA);\cite{LG1968} likewise the Cu--N separation of 1.887(10)\,\AA\ is comparable to that found in CuCN ($d($Cu--C/N$)=1.839(9)$--$1.872(12)$\,\AA).\cite{HEC2004}

\begin{table}[b]
\small
  \caption{\ Structural details for Cu$_3$[Co(CN)$_6$] obtained by Pawley/Rietveld refinement against X-ray powder diffraction data collected at 300\,K and estimated 0\,K values extracted from linear fits to 100--598\,K refinements. Atom positions are Co $(0,0,0)$, Ag $(\frac{1}{2},0,\frac{1}{2})$, C $(x_{\textrm C},0,z_{\textrm C})$, N $(x_{\textrm N},0,z_{\textrm N})$.}
  \label{table1}
  \begin{tabular*}{0.48\textwidth}{@{\extracolsep{\fill}}lll}
    \hline
     & 300\,K (experimental) & 0\,K (estimated) \\
    \hline
    Crystal system&Trigonal&Trigonal\\
    Space group&$P\bar31m$&$P\bar31m$\\
    $a$ (\AA)&6.9085(10)&6.8552\\
    $c$ (\AA)&6.7077(16)&6.7970\\
    $V$ (\AA$^3$)&277.25(8)&276.66\\
    $x_{\textrm C}$&0.2177(15)&0.2167\\
    $z_{\textrm C}$&0.1566(14)&0.1533\\
    $x_{\textrm N}$&0.3161(15)&0.3182\\
    $z_{\textrm N}$&0.2920(14)&0.2887\\
    $B_{\textrm{iso}}$ (\AA$^2$)&3.91(14)&--\\
    \hline
  \end{tabular*}
\end{table}

A property of the particular space group symmetry of Cu$_3$[Co(CN)$_6$] is that the Cu$\ldots$Cu separation is directly related to the lattice parameters:
\begin{equation}
r_{\textrm{Cu\ldots Cu}}=\frac{a}{2}.
\end{equation}
Hence we find $r_{\textrm{Cu\ldots Cu}}=3.4543(5)$\,\AA, which lies at the very upper bound of Cu$\ldots$Cu separations for which cuprophilic interactions are considered relevant.\cite{KSL2008} One crude measure of the strength of metallophilic interactions is to consider the ratio of the observed interatomic distance to the sum of the corresponding vdW radii.\cite{Jansen_1987} Using our room-temperature lattice parameters and the vdW radii given in Ref.~\citenum{B2001} we obtain a ratio of 1.00 for Cu$_3$[Co(CN)$_6$], which is remarkably similar to the corresponding value for Ag$_3$[Co(CN)$_6$] (0.99).\cite{GCC2008} So, at face value, one might expect comparable metallophilic interaction strengths for the two systems.


\subsection{Thermal expansion behaviour}

\begin{figure}
\centering
  \includegraphics{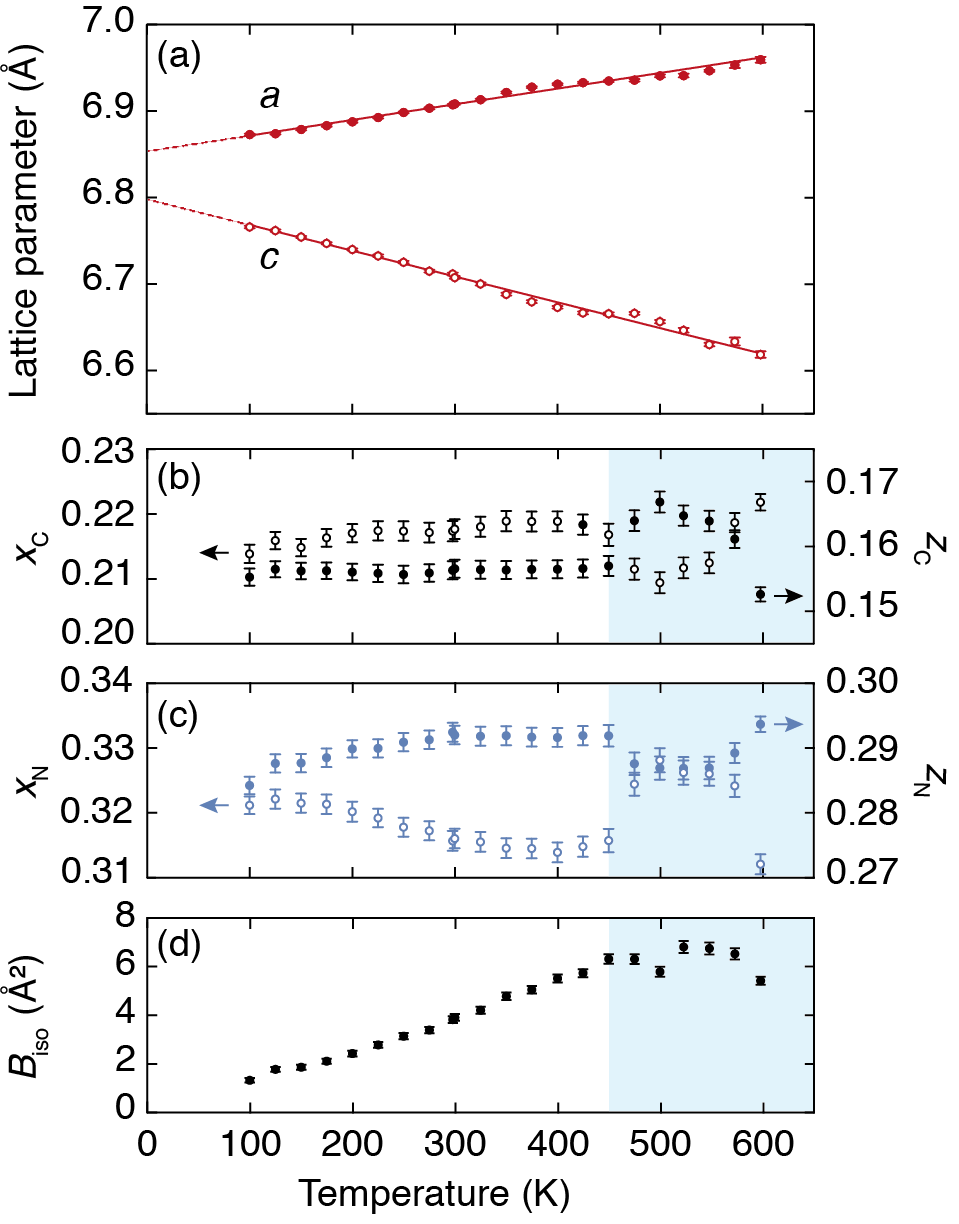}
  \caption{Temperature dependence of structural parameters of Cu$_3$[Co(CN)$_6$] as determined using variable-temperature X-ray powder diffraction. (a) Lattice parameters $a$ and $c$ (filled and open symbols, respectively), together with the linear fits (solid lines) used to determine the uniaxial coefficients of thermal expansion. The fits are extrapolated to 0\,K (dashed lines) to give the corresponding `0\,K estimates' discussed in the text. (b,c) Positional coordinates for the C and N atoms, showing smooth variation over the temperature range 100--450\,K for which reliable Rietveld refinements were obtained. The temperature regime 450--600\,K is shaded as refinements in this regime gave reliable lattice parameters but unreliable positional coordinates and atomic displacement parameters. (d) Isotropic atomic displacement parameter $B_{\textrm{iso}}=8\pi^2\langle u\rangle^2$ used to model thermal displacements for all atoms.}
  \label{fig4}
\end{figure}

Having collected a series of X-ray diffraction patterns over the temperature range 100--598\,K, we carried out Rietveld refinements for each data set using the same approach described above. We obtained satisfactory refinements for all temperatures, albeit with some signs of increased uncertainties at the very highest temperatures---\emph{i.e.}\ close to the onset of decomposition of the Prussian blue phase. The temperature dependence of the lattice parameters, illustrated in Fig.~\ref{fig4}(a), was observed to be approximately linear over this entire temperature range. As in nearly all members of this structural family, Cu$_3$[Co(CN)$_6$] exhibits an NTE effect parallel to the $c$ crystal axis, and PTE effects in perpendicular directions (\emph{i.e.}, including the $a$ and $b$ crystal axes). Hence the basic thermomechanical response of this system can be understood in terms of the same `wine-rack' mechanism illustrated in Fig.~\ref{fig1}. The remaining structural parameters $x_{\textrm C},z_{\textrm C},x_{\textrm N},z_{\textrm N},B_{\textrm{iso}}$ also show linear temperature dependencies [Fig.~\ref{fig4}(b--d)]; taken together these values allow us to estimate a set of 0\,K structural parameters that may prove useful for comparison against \emph{e.g.}\ \emph{ab initio} studies [Table~\ref{table1}].

Coefficients of thermal expansion were determined using linear fits to the lattice parameter data,\cite{pascal} and are given in Table~\ref{table2}. What is immediately apparent is that the magnitudes of both PTE and NTE effects in Cu$_3$[Co(CN)$_6$] are substantially smaller than those in the Ag-containing system. Consequently, Cu$_3$[Co(CN)$_6$] is not a colossal thermal expansion material, and its thermomechanical response shares more in common with other Cu-containing networks such as $\alpha$-Cu[C(CN)$_3$] (Ref.~\citenum{Hunt_2015}) and CuCN (Ref.~\citenum{HWB2010}) than with Ag$_3$[Co(CN)$_6$] and Ag$_3$[Fe(CN)$_6$].\cite{GKT2008a} We will come to rationalise the differences in behaviour of the copper(I) and silver(I) hexacyanocobaltates below, but include first some additional analysis of the trends in lattice parameters we observe using our newly-measured data.

\begin{table}[b]
\small
  \caption{\ Experimental coefficients of thermal expansion for A$_{\textrm3}$[Co(CN)$_{\textrm6}$] systems.}
  \label{table2}
  \begin{tabular*}{0.48\textwidth}{@{\extracolsep{\fill}}llllll}
    \hline
    A & $\alpha_a$& $\alpha_c$ & $\alpha_V$ & $\Delta T$ &Ref. \\
     & (MK$^{-1}$)& (MK$^{-1}$)&  (MK$^{-1}$) &(K)\\
    \hline
    H&12.0(4)&$-$8.8(3)&15.1(6)&4--300&\citenum{KDE2010}\\
    Cu&25.4(5)&$-$43.5(8)&7.4(11)&100--598&This work\\
    Ag&145.9(6)&$-$122.1(3)&169.8(9)&16--500&\citenum{GCC2008}\\
    \hline
  \end{tabular*}
\end{table}

The `wine-rack' mechanism that is thermally activated in this system can be interrogated directly using the so-called mechanical building unit (XBU) approach.\cite{OCM2012} We make use of the pair of transformations
\begin{eqnarray}
r&=&\frac{1}{2}\sqrt{a^2+c^2},\\
\theta&=&\tan^{-1}\left(\frac{c}{a}\right),
\end{eqnarray}
which relate the unit cell dimensions to the framework strut length $r$ and framework angle $\theta$ [Fig.~\ref{fig3}]. Using these same relationships, we can recast the lattice expansivities in terms of XBU expansivities, obtaining $\alpha_r=-8.2$\,MK$^{-1}$ and $\alpha_{\theta}=43.1$\,MK$^{-1}$. Hence the bulk of the thermal expansion response can be accounted for by changes in the framework geometry ($|\alpha_{\theta}|\gg|\alpha_r|$); the lattice expansivities attributable to this flexing mechanism alone are $\alpha_a^\prime=33.5$\,MK$^{-1}$ and $\alpha_c^\prime=-35.7$\,MK$^{-1}$, where we use the prime notation to indicate calculation from $\alpha_{\theta}$. The observation $\alpha_r<0$ indicates that the Co--CN--Cu--NC--Co `struts' from which the framework structure of Cu$_3$[Co(CN)$_6$] is assembled actually contract with increasing temperature. This behaviour is likely due to thermal activation of transverse vibrational modes where lateral displacements of the chain (maximal at the Cu site) require shortening of the Co$\ldots$Co vector.\cite{Mary_1996,Barrera_2005} Such a mechanism is implicated in the uniaxial NTE of CuCN itself ($\alpha_{\textrm{chain}}=-32.1$\,MK$^{-1}$, Ref.~\citenum{HWB2010,HEC2004}), and is presumably tempered here somewhat relative to that system by the increased strength of Co$^{\textrm{III}}$--C \emph{vs} Cu$^{\textrm I}$--C bonds.\cite{Sharpe_1976}


One consequence of the negative value of $\alpha_r$ is that the volume coefficient of thermal expansion of Cu$_3$[Co(CN)$_6$] is unusually small for systems in this particular family. Formally, this situation arises because of the fortuitous equivalence $\alpha_r\simeq-\frac{1}{3}|\alpha^\prime|$, which is the geometric requirement for $\alpha_V\rightarrow0$.\footnote{Note that $\alpha_i=\alpha_i^\prime+\alpha_r$, and hence $\alpha_V\sim\alpha_a^\prime+3\alpha_r$.} Hence this material has the unusual property of (approximately) temperature-independent density despite its relatively large linear thermal expansivities. At face value, this property may be expected to result in unusually extreme uniaxial compressibilities under application of hydrostatic pressure, since small changes in volume would appear to be linked to large changes in lattice dimensions. However, we anticipate by analogy to related systems that the XBU compressibility $K_r$ is actually positive rather than negative, and so a small $\alpha_V$ need not require a large bulk modulus.\cite{Adamson_2015,OCM2012} Nevertheless we expect the particular uniaxial compressibility corresponding to the $c$ crystal axis to be negative, and so investigation of the NLC behaviour of Cu$_3$[Co(CN)$_6$] could prove a fruitful avenue of future research.

\subsection{\emph{Ab initio} calculations}

The observation of more moderate thermal expansion behaviour in Cu$_3$[Co(CN)$_6$] relative to that in Ag$_3$[Co(CN)$_6$] poses a simple question: does this situation arise because cuprophilic interactions are actually \emph{stronger} than argentophilic interactions, and hence less susceptible to changes in temperature?

\begin{figure*}
\centering
  \includegraphics{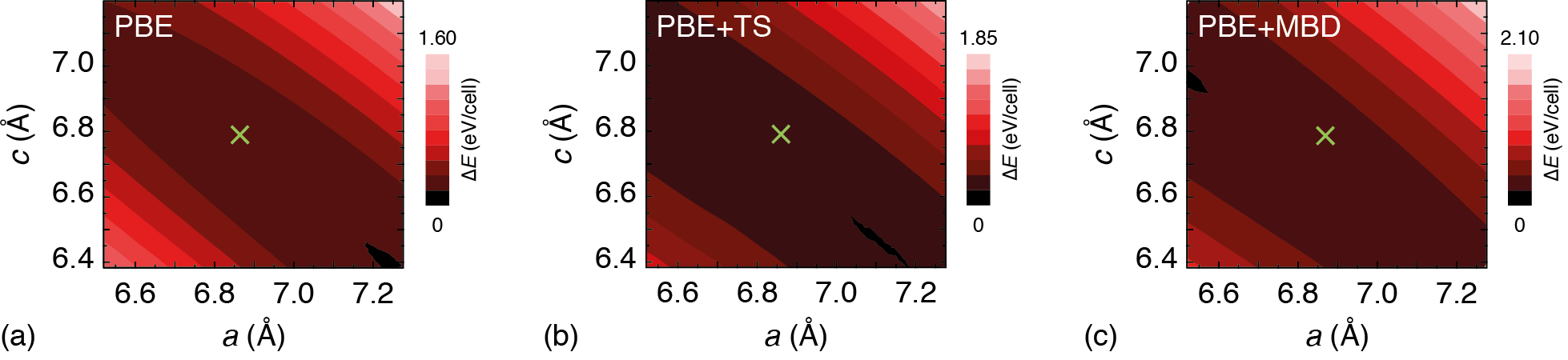}
  \caption{The (a) PBE, (b) PBE+TS, and (c) PBE+MBD potential energy surfaces of Cu$_3$[Co(CN)$_6$] as a function of unit cell dimensions. The experimental lattice constants are indicated by crosses. Energies are given relative to the ground state in each case.}
  \label{figpes}
\end{figure*}

In order to answer this question, we turn to \emph{ab initio} calculations, which if carried out so as to include consideration of vdW interactions allow direct quantification of the metallophilic interactions in both compounds. We begin by reporting the 0\,K structure for Cu$_3$[Co(CN)$_6$] obtained computationally and demonstrate that the inclusion of dispersive interactions is necessary to obtain good consistency with our experimental results. By mapping out the potential energy surface (PES) for all three A$_3$[Co(CN)$_6$] systems (A = H, Cu, Ag) across a variety of lattice strains and then taking into account the variation in vdW energies at each point, we extract the free-atom and in-solid (effective) $C_6$ coefficients. The value of these coefficients for each atom type A acts as a measure of the strength of metallophilic interactions in the corresponding A$_3$[Co(CN)$_6$] system.

\begin{table}[b]
\small
  \caption{Comparison between experimental and \emph{ab initio} lattice parameters for Cu[Co(CN)$_6$]. The difference term $\Delta$ corresponds to the sum of absolute cell strains $\sum_i|(x_{i,\textrm{calc}}-x_{i,\textrm{exp}})/x_{i,\textrm{exp}}|$.}
  \label{tableucps}
  \begin{tabular*}{0.48\textwidth}{@{\extracolsep{\fill}}lllll}
    \hline
    &exp. (0\,K)&PBE&TS&MBD\\
    \hline
    $a$ (\AA)&6.8552&7.267&7.130&6.495\\
    $c$ (\AA)&6.7970&6.365&6.432&6.978\\
    $V$ (\AA$^3$)&276.62&291.06&283.00&254.98\\
    $\Delta$ (\%)&0&18.4&13.4&13.2\\
    \hline
  \end{tabular*}
\end{table}

\begin{figure}[b]
\centering
  \includegraphics{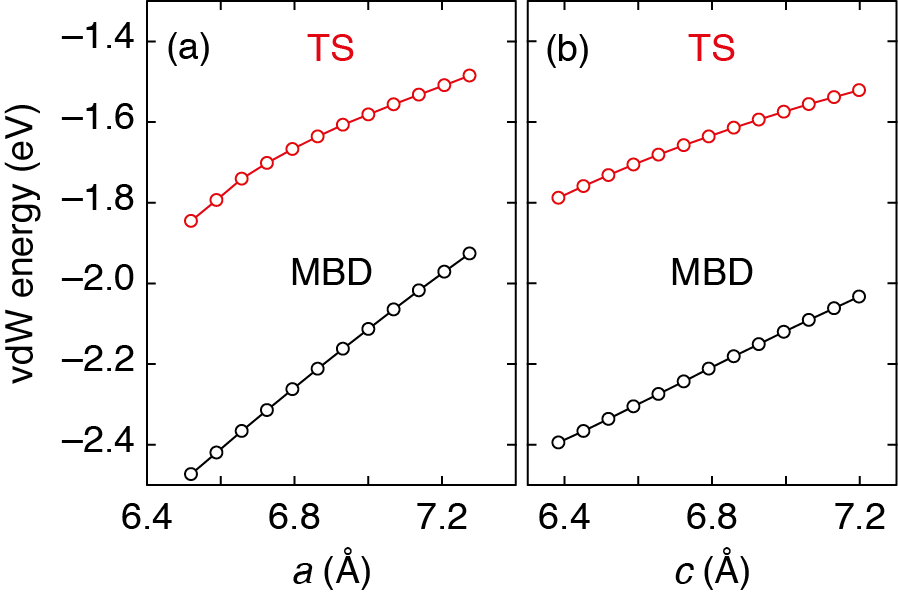}
  \caption{The TS and MBD vdW energies in Cu$_3$[Co(CN)$_6$] per unit cell (a) as a function of lattice constant $a$ with $c$ fixed to experimental values and (b) as a function of $c$ with $a$ fixed to experimental values.}
  \label{figvdwplot}
\end{figure}

The unit cell dimensions obtained in our DFT + vdW calculations are given in Table~\ref{tableucps}. The influence of dispersion energy on the lattice constants is large, just as is now known to be the case for Ag$_3$[Co(CN)$_6$].\cite{Liu_2016} Our PBE calculation without vdW interactions overestimates $a$ and underestimates $c$. Upon including dispersion interactions the lattice constants move closer to the experimental values. We note that the enhanced cohesive MBD energy for Cu$_3$[Co(CN)$_6$] arises from the collective effect of vdW interactions and the self-consistent polarisation in the unit cell.\cite{Liu_2016} In Figure~\ref{figpes} we show a representative section of the PES for the three calculation regimes, and Figure~\ref{figvdwplot} shows the TS and MBD vdW energies as a function of the individual $a$ and $c$ lattice constants. Our results make clear that the vdW energy depends more strongly on $a$ than it does on $c$. Since the framework strut length $r$ is more rigid than the framework angle $\theta$, then to lower the total energy the lattice simply contracts along $a$ (and $b$) while expanding along $c$. Hence the same mechanism explains the qualitative change in lattice constants observed both as a result of using different vdW calculation methods and as a result of an increase in the polarisability of atom A. Indeed because the MBD energy depends almost linearly on the lattice constants it behaves as an effective pressure on the lattice, equivalent to 1.22\,GPa along $a$ and 1.76\,GPa along $c$.

To compare the strength of cuprophilic interactions in Cu$_3$[Co(CN)$_6$] with that of argentophilic interactions in Ag$_3$[Co(CN)$_6$] we further analysed our DFT+vdW results. Our basic approach was to parameterise the vdW contribution to the TS-vdW energy in terms of dispersion coefficients $C_6$ and vdW radii $R_0$ for each atom type. In the PBE+TS calculations, the free-atom $C_6$ coefficient and vdW radii $R_0$ are used as the initial input parameters. The effect of the local chemical environment is taken into account by calculating the effective in-solid $C_6$ and $R_0$ as described in Ref.~\citenum{Tkatchenko_2009}. Table~\ref{tablevdw} lists our results for the free-atom vdW parameters and the effective parameters for A$_3$[Co(CN)$_6$] (A = Ag, Cu, H) at the experimental lattice constants. We find that the argentophilic interactions are indeed stronger than cuprophilic interactions in these systems, as both the free-atom and effective $C_6$ values are larger by $\sim40$\% for Ag relative to Cu. For completeness we note that the effect of the local chemical environment on the $C_6$ coefficients is to reduce the dispersion coefficients.


\begin{table}
\small
  \caption{The PBE+TS free-atom and in-solid vdW parameters for A atoms in A$_3$[Co(CN)$_6$] (A = Ag, Cu, H) at experimental lattice constants.}
  \label{tablevdw}
  \begin{tabular*}{0.48\textwidth}{@{\extracolsep{\fill}}lllll}
    \hline
    & \multicolumn{2}{c}{$C_6$ (hartree\,bohr$^6$)} & \multicolumn{2}{c}{$R_0$ (bohr)}\\
    &free-atom&in-solid&free-atom&in-solid\\
    \hline
    Ag$_3$[Co(CN)$_6$]&339.00&295.73&3.82&3.73\\
    Cu$_3$[Co(CN)$_6$]&235.00&207.03&3.76&3.64\\
    H$_3$[Co(CN)$_6$]&6.50&4.28&3.10&2.89\\
    \hline
  \end{tabular*}
\end{table}

\subsection{Lattice dynamical calculations}

We supplement these high-level \emph{ab initio} results with a series of extremely simple lattice-dynamical calculations that also allow us to estimate the relative strengths of metallophilic interactions in Cu$_3$[Co(CN)$_6$] and Ag$_3$[Co(CN)$_6$]. The approach we use is to develop the very simplest abstraction of all three A$_3$[Co(CN)$_6$] systems (A = H, Cu, Ag) that captures the key interactions responsible for their thermomechanical response. We parameterise this model with sufficiently few variables that six experimental observables (the two independent lattice parameters for each of the three systems) can be used to estimate the relative magnitudes of the metallophilic interactions in the A = Cu, Ag compounds.

\begin{figure}[b]
\centering
  \includegraphics{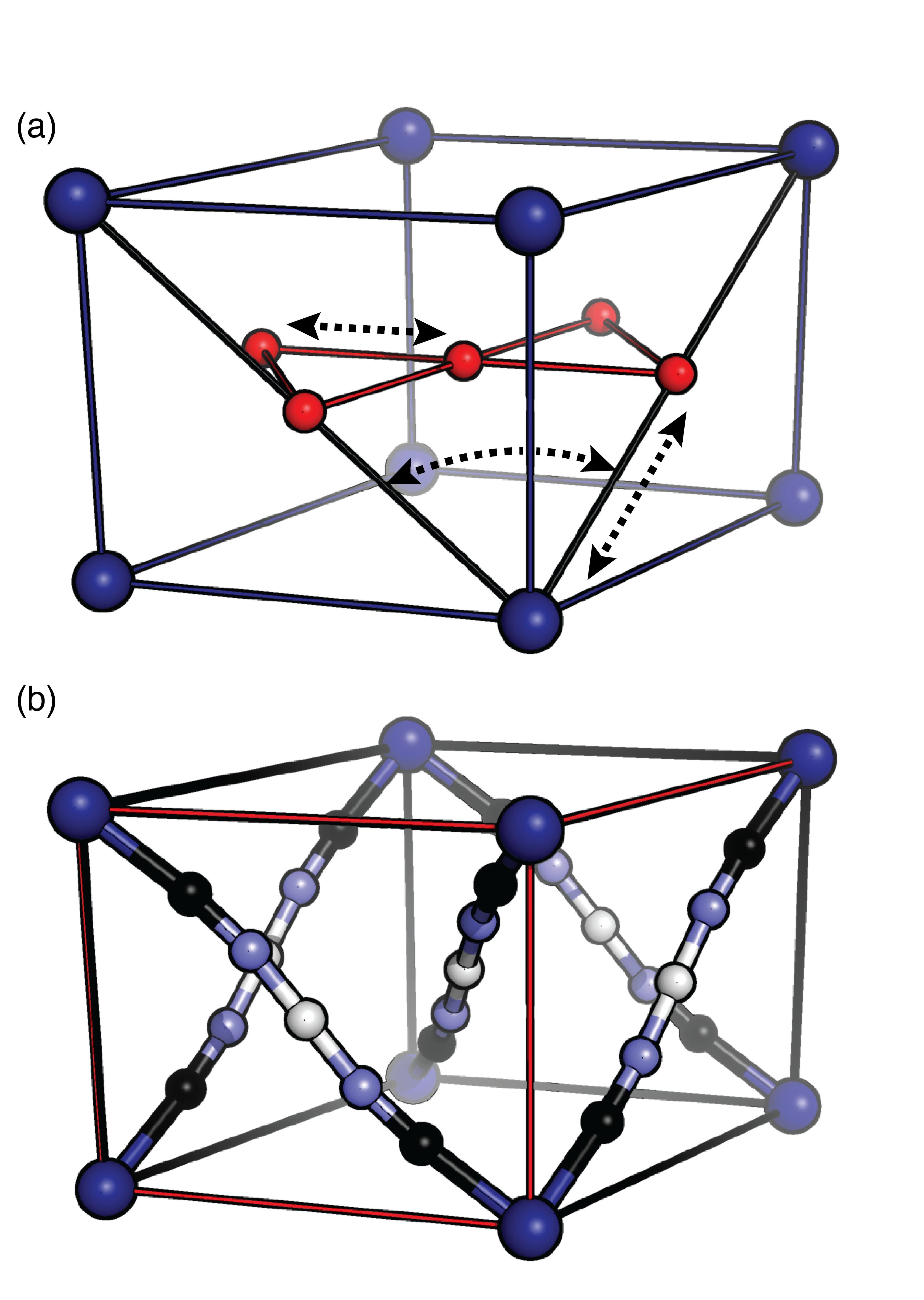}
  \caption{Lattice dynamical model for A$_3$[Co(CN)$_6$] systems and the corresponding match in H$_3$[Co(CN)$_6$] geometry used to estimate its interaction potential parameters. (a) The model consists of X atoms at the Co site (large blue spheres; formal charge $-1.5e$) and A atoms at the H/Cu/Ag site (red spheres; formal charge $+0.5e$). The model includes three interatomic potentials in addition to Coulomb interactions: harmonic Co--A `bond stretching' interactions, harmonic A--Co--A `bond bending' interactions, and $r^{-6}$ dispersive interactions between A sites. (b) Match between experimental unit cell dimensions (solid black lines) of H$_3$[Co(CN)$_6$] (Ref.~\citenum{KDE2010}) and relaxed cell in our lattice dynamical model (solid red lines) for the parameter values given in Table~\ref{table3}.}
  \label{fig5}
\end{figure}

The same structural model is used for all three systems: $P\bar31m$ crystal symmetry, with a single anion (mass $m=m({\textrm{Co}})+6m({\textrm{C}})+6m({\textrm{N}})$) of charge $-1.5e$ at position $(0,0,0)$ and a cation ($m=m(\textrm{A})$) with charge $+0.5e$ at position $(\frac{1}{2},0,\frac{1}{2})$ [Fig.~\ref{fig5}(a)]. These charges reflect the approximate Mulliken charges determined for H$_3$[Co(CN)$_6$] and Ag$_3$[Co(CN)$_6$] in Ref.~\citenum{CGD2008} and are consistent with the Hirshfeld and Bader charges obtained in our own \emph{ab initio} calculations (see SI). We refer to the anion using the symbol X (formally this corresponds to the [Co(CN)$_6$]$^{3-}$ ion), giving the unit cell composition A$_3$X. This structural model is then decorated with three interaction potentials: first, a harmonic bond potential between neighbouring A and X sites
\begin{equation}
E_{\textrm{harm}}=\frac{1}{2}k_{\textrm{harm}}(r_{\textrm{A--X}}-r_0)^2;
\end{equation}
second, a harmonic bond angle potential governing A--X--A triplets
\begin{equation}
E_{\textrm{angle}}=\frac{1}{2}k_{\textrm{angle}}(\theta_{\textrm{A--X--A}}-\theta_0)^2;
\end{equation}
and, third, (in the case of Cu and Ag systems) dispersive interactions between neighbouring A sites intended to reflect the empirical $\frac{1}{r^6}$-dependence of metallophilic interactions\cite{MSR2002}
\begin{equation}
E_{\textrm{disp}}=-\frac{C_6}{(r_{\textrm{A\ldots A}})^6}.
\end{equation}
In this way the lattice energy is given by the sum
\begin{equation}
E_{\textrm{latt}}=E_{\textrm{Coulomb}}+E_{\textrm{harm}}+E_{\textrm{angle}}+E_{\textrm{disp}}.
\end{equation}

In order to reduce the number of parameters involved in this model, we make the following assumptions. First, we take the effective charges at X and A sites to be system-independent. We justify this assumption by noting that the Mulliken charges reported for H$_3$[Co(CN)$_6$] and Ag$_3$[Co(CN)$_6$] vary more greatly by calculation method than they do between systems;\cite{CGD2008} the A = Cu case is intermediate to the A = H and A = Ag cases (see SI). Second, we take the flexing stiffness $k_{\textrm{angle}}$ and equilbrium angle $\theta_0$ also to be system-independent. This is probably reasonable given that both terms will be governed by the chemistry of the [Co(CN)$_6$]$^{3-}$ anion, which is common to all three systems. Third, we take the (system-dependent) values of $r_0$ as the sum of bond lengths $d($Co--C$)+d($C--N$)+d($N--A$)$ determined crystallographically: we use the values from Ref.~\citenum{KDE2010} for A = H, from Ref.~\citenum{GKT2008b} for A = Ag, and from our present study for A = Cu. Fourth, we assume that the $E_{\textrm{disp}}$ term is negligible for the A = H case, which is likely given the small electron density expected at the A site for this system (and is certainly supported by our own \emph{ab initio} results as given in Table~\ref{tablevdw}).

Our starting point is to determine a set of parameters $k_{\textrm{harm}}, k_{\textrm{angle}}, \theta_0$ that, when used to drive a lattice-dynamical geometry optimisation, result in the closest possible agreement between 0\,K (derived from experiment) and relaxed cell parameters for A = H. We make the additional requirement that $\theta_0$ should be as close to $90^\circ$ as possible. The parameter set we obtain is listed in Table~\ref{table3}, together with a comparison of the experimental and simulated lattice parameters; the corresponding match in framework geometry is illustrated in Fig.~\ref{fig5}(b).\footnote{We found the quality of fit was relatively insensitive to changes in $k_{\textrm{harm}}$ of up to \emph{ca} 25\% of its value.  Variations in this parameter did affect the absolute values of the compressibilities determined subsequently; however the same trend in magnitudes of compressibilities shown in Fig.~\ref{fig6} was found in all cases.} We note that we do not attach any particular physical meaning to the parameter values in our model, since (in particular) the charge distribution we use is a heavily simplified representation of reality. Nevertheless it is reassuring that even this simple model allows robust geometry optimisation to a physically-sensible state.

\begin{table}
\small
  \caption{\ Lattice dynamical model parameters and comparison between calculated and observed lattice parameters. Refined parameters are shown in bold.}
  \label{table3}
  \begin{tabular*}{0.48\textwidth}{@{\extracolsep{\fill}}llll}
    \hline
    &H$_3$[Co(CN)$_6$]&Cu$_3$[Co(CN)$_6$]&Ag$_3$[Co(CN)$_6$]\\
    \hline
    $k_{\textrm{harm}}$ (eV/\AA$^2$)&{\bf 400}&400&400\\
    $r_0$ (\AA)&4.319&4.867&5.070\\
    $k_{\textrm{angle}}$ (eV/rad$^2$)&{\bf 47}&47&47\\
    $\theta_0$ ($^\circ$)&{\bf 89}&89&89\\
    $C_6$ (eV\,\AA$^6$)&0&{\bf 8810}&{\bf 14400}\\
    \hline
    $a$ (\AA)&6.450&6.901&6.812\\
    $a_{\textrm{expt}}^{0\,\textrm K}$ (\AA)&6.409&6.855&6.740\\
    $\Delta a/a$ (\%)&+0.6\%&+0.7\%&+1.1\%\\
    $c$ (\AA)&5.749&6.842&7.474\\
    $c_{\textrm{expt}}^{0\,\textrm K}$ (\AA) &5.713&6.797&7.390\\
    $\Delta c/c$ (\%)&+0.6\%&+0.7\%&+1.1\%\\
    \hline
  \end{tabular*}
\end{table}

Having used the geometry of the A = H system to determine all of the system-independent parameter values, we proceeded to optimise the geometry of analogous models for A = Cu and Ag. In each case the value of $r_0$ was updated according to the experimental bond lengths, and only the value of $C_6$ was varied in order to obtain the closest match between calculated and experimental (0\,K extrapolated) lattice parameters. The corresponding parameter values and optimised cell dimensions are again summarised in Table~\ref{table3}; we note that the level of agreement ($<2\%$) is encouraging given the simplicity of the lattice dynamical model we have used. Also encouraging is that, for both compounds, the $a$ lattice parameters are overestimated in the absence of a metallophilic contribution to the lattice enthalpy. This mirrors the results of vastly higher-level \emph{ab initio} geometry optimisations,\cite{CGD2008,Fang_2014} and indicates that the electrostatic contribution to the free energy (the single component of our model acting to increase $a$) operates in tension with the metallophilic interactions. While we do not attach any importance to the absolute values of the $C_6$ parameters that emerge from our calculations, what we do think is meaningful is the observation that $C_6$ is larger for A = Ag than for A = Cu. In other words, the experimental unit cell dimensions for Cu$_3$[Co(CN)$_6$] and Ag$_3$[Co(CN)$_6$] are consistent with stronger argentophilic interactions in the latter than cuprophilic interactions in the former. Moreover, the ratio of cuprophilic:argentophilic interaction strengths we deduce from our simple lattice-dynamical model is essentially the same as that obtained in our \emph{ab initio} calculations: $C_6$(Cu)/$C_6$(Ag) = 61\% \emph{vs} 70\%, respectively. 

\subsection{Flexibility from competing interactions}

So our various calculations converge on the same scenario whereby cuprophilic interactions in Cu$_3$[Co(CN)$_6$] are weaker than argentophilic interactions in Ag$_3$[Co(CN)$_6$] by 30--40\%. One obvious question remains: how is this observation consistent with the more moderate thermal expansion behaviour of the Cu-containing compound?

\begin{figure}
\centering
  \includegraphics{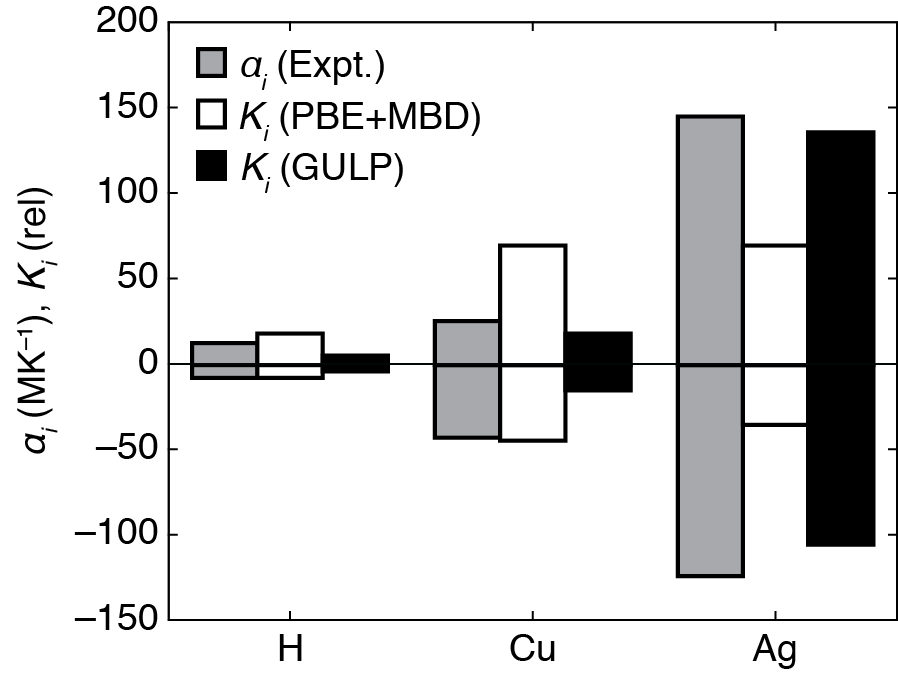}
  \caption{Trends in calculated uniaxial compressibilities (white bars = \emph{ab initio}; black bars = lattice-dynamical; data normalised for comparison) and lattice expansivities (grey bars = values taken from Refs.~\citenum{GCC2008, KDE2010} and this study) for A$_3$[Co(CN)$_6$] compounds.}
  \label{fig6}
\end{figure}

To address this question we exploit the approximate proportionality between thermal expansivities and isothermal compressibilities noted in Refs.~\citenum{Munn_1972,GKT2008b,Cairns_2015}:
\begin{equation}
\alpha_i\simeq\frac{C_T}{V}\hat\gamma K_i.
\end{equation}
Here $C_T$ is the isothermal specific heat, $V$ the molar volume, $\hat\gamma$ the mean effective Gr{\"u}neisen parameter and $K_i$ the uniaxial compressibilities. We estimate that the pre-factor $C_T\hat\gamma/V$ varies by not more than $\sim25\%$ between the A = Cu and A = Ag systems,\footnote{Here we have made use of three relationships: first, that $\hat\gamma$ appears to be relatively system-independent;\cite{CGD2008} second, that the ratio of the $C_T$ values for A = Cu and Ag will be approximately equal to the ratio of the $\sqrt{m}$ terms, since the low-energy phonon dispersion will be dominated by heavy-atom displacements; and third, we use the experimental molar volumes.} such that a comparison of compressibilities for the two compounds provides a reasonable first-order approximation to the relative thermal expansivities. We concern ourselves with compressibilities rather than expansivities since the former are obtainable directly from the calculations (both \emph{ab initio} and lattice-dynamical) described above. The relative compressibilities for all three compounds are illustrated graphically in Fig.~\ref{fig6}. What is evident is that the Cu-containing compound exhibits intermediate behaviour to the H- and Ag-containing systems, despite its relatively weaker metallophilic interactions. The qualitative similarity to the relative thermal expansivities is striking, particularly given the (necessary) omission of anharmonic contributions from our calculations which likely contribute substantially to the experimental behaviour.\cite{Dove_2016}

\section{Concluding remarks}

\begin{figure}
\centering
  \includegraphics{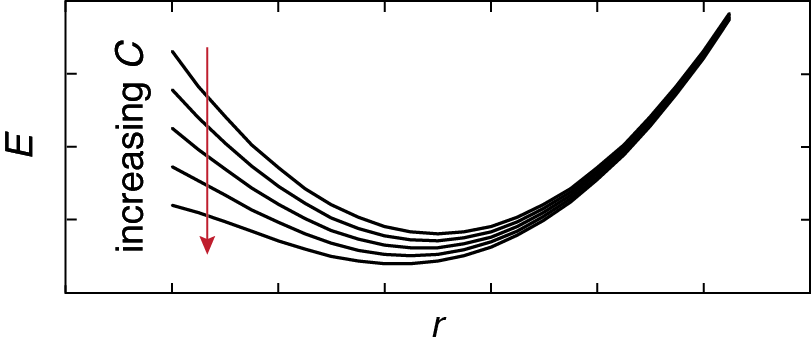}
  \caption{Flattening of an effective interaction potential $E=\frac{1}{2}k(r-r_0)^2+C_6r^{-6}$ with increasing dispersion interaction strength $C_6$. Reduced curvature leads to more extreme expansivity and compressibility behaviour.}
  \label{fig7}
\end{figure}

We are led to the counterintuitive conclusion that stronger interactions can actually make a material more compliant: Ag$_3$[Co(CN)$_6$] exhibits colossal thermomechanical responses but Cu$_3$[Co(CN)$_6$] does not, despite the energy scale associated with metallophilic interactions being larger in the former than in the latter. Of course the key here is that metallophilic interactions are net attractive, and act in tension with the (repulsive) electrostatic component.\cite{Fang_2014,Hermet_2013} Any effective harmonic potential can be made increasingly shallow by the addition of attractive $r^{-6}$ terms, as illustrated in Fig.~\ref{fig7}. This is the nub of the physics at play in this family: in the absence of metallophilic interactions, the frameworks are not especially mechanically responsive but they do become so as metallophilicity is introduced. 

Hence the conventional materials design rules are reversed, and we anticipate that the member of the A$_3$[Co(CN)$_6$] family likely to show the most extreme thermomechanical response is actually the as-yet-unrealised compound Au$_3$[Co(CN)$_6$]. It was shown in Ref.~\citenum{CGD2008} that this system is likely to have a particularly compliant structure, although the degree of compliance will depend heavily on the strength of the aurophilic interaction contribution to the lattice enthalpy. Given the notorious difficulty of accessing aqueous Au(I) chemistry, it is not yet clear how Au$_3$[Co(CN)$_6$] might be accessed synthetically. A viable alternative is the (also unrealised) compound Fe[Au(CN)$_2$]$_3$---\emph{i.e.}, with Co(III) replaced by Fe(III) and the CN ion orientations reversed---which by analogy to Fe[Ag(CN)$_2$]$_3$ should in principle be accessible \emph{via} reaction of aqueous Fe$^{3+}$-containing solutions with KAu(CN)$_2$.\cite{Hill_2015} The observation\cite{GKT2008a} of qualitatively similar `colossal' thermal expansion in Ag$_3$[Co(CN)$_6$] and Ag$_3$[Fe(CN)$_6$] suggests that chemical substitution at the trivalent metal site is unlikely to influence the degree of thermomechanical response observed.

From a computational perspective, one key implication of our study is the importance of obtaining accurate descriptions of vdW interactions in compliant framework materials. This importance is particularly acute for systems such as Cu$_3$[Co(CN)$_6$] and Ag$_3$[Co(CN)$_6$] where the PES is anomalously shallow as a result of competition between vdW and electrostatic contributions. A key challenge in this regard is the treatment of finite-temperature effects; \emph{i.e.}\ anharmonicity. We anticipate that the discovery of anomalous mechanics in increasingly many systems based on vdW-type interactions\cite{Cairns_2013,Woodall_2013} will motivate further research effort along precisely these lines.


\section*{Acknowledgements}

The synchrotron diffraction measurements were carried out at the Diamond Light Source (I11 Beamline). We are extremely grateful for the award of a Block Allocation Grant, which made this work possible, and for the assistance in data collection provided by S. J. Cassidy (Oxford) and the I11 beamline staff. A.F.S., A.R.O. and A.L.G. gratefully acknowledge funding through the European Research Council (Grant 279705). C.S.C. and A.L.G. acknowledge funding through the Leverhulme Trust (Grant RPG-2015-292). A.R.O. thanks the Diamond Light Source for a Studentship. This project received funding from the European Union (EU) Horizon 2020 Research and Innovation Programme under Marie Sklodowska-Curie Grant Agreement 641887 (project acronym DEFNET).



\balance


\bibliography{rsc} 

\providecommand*{\mcitethebibliography}{\thebibliography}
\csname @ifundefined\endcsname{endmcitethebibliography}
{\let\endmcitethebibliography\endthebibliography}{}
\begin{mcitethebibliography}{65}
\providecommand*{\natexlab}[1]{#1}
\providecommand*{\mciteSetBstSublistMode}[1]{}
\providecommand*{\mciteSetBstMaxWidthForm}[2]{}
\providecommand*{\mciteBstWouldAddEndPuncttrue}
  {\def\EndOfBibitem{\unskip.}}
\providecommand*{\mciteBstWouldAddEndPunctfalse}
  {\let\EndOfBibitem\relax}
\providecommand*{\mciteSetBstMidEndSepPunct}[3]{}
\providecommand*{\mciteSetBstSublistLabelBeginEnd}[3]{}
\providecommand*{\EndOfBibitem}{}
\mciteSetBstSublistMode{f}
\mciteSetBstMaxWidthForm{subitem}
{(\emph{\alph{mcitesubitemcount}})}
\mciteSetBstSublistLabelBeginEnd{\mcitemaxwidthsubitemform\space}
{\relax}{\relax}

\bibitem[Coudert(2015)]{Coudert_2015}
F.-X. Coudert, \emph{Chem. Mater.}, 2015, \textbf{27}, 1905--1916\relax
\mciteBstWouldAddEndPuncttrue
\mciteSetBstMidEndSepPunct{\mcitedefaultmidpunct}
{\mcitedefaultendpunct}{\mcitedefaultseppunct}\relax
\EndOfBibitem
\bibitem[Ogborn \emph{et~al.}(2012)Ogborn, Collings, Moggach, Thompson, and
  Goodwin]{OCM2012}
J.~M. Ogborn, I.~E. Collings, S.~A. Moggach, A.~L. Thompson and A.~L. Goodwin,
  \emph{Chem. Sci.}, 2012, \textbf{3}, 3011--3017\relax
\mciteBstWouldAddEndPuncttrue
\mciteSetBstMidEndSepPunct{\mcitedefaultmidpunct}
{\mcitedefaultendpunct}{\mcitedefaultseppunct}\relax
\EndOfBibitem
\bibitem[Horike \emph{et~al.}(2009)Horike, Shimomura, and
  Kitagawa]{Horike_2009}
S.~Horike, S.~Shimomura and S.~Kitagawa, \emph{Nature Chem.}, 2009, \textbf{1},
  695--704\relax
\mciteBstWouldAddEndPuncttrue
\mciteSetBstMidEndSepPunct{\mcitedefaultmidpunct}
{\mcitedefaultendpunct}{\mcitedefaultseppunct}\relax
\EndOfBibitem
\bibitem[Krishna \emph{et~al.}(2016)Krishna, Devarapalli, Lal, and
  Reddy]{Krishna_2016}
G.~R. Krishna, R.~Devarapalli, G.~Lal and C.~M. Reddy, \emph{J. Am. Chem.
  Soc.}, 2016, \textbf{138}, 13561--13567\relax
\mciteBstWouldAddEndPuncttrue
\mciteSetBstMidEndSepPunct{\mcitedefaultmidpunct}
{\mcitedefaultendpunct}{\mcitedefaultseppunct}\relax
\EndOfBibitem
\bibitem[Das \emph{et~al.}(2010)Das, Jacobs, and Barbour]{Das_2010}
D.~Das, T.~Jacobs and L.~J. Barbour, \emph{Nature Mater.}, 2010, \textbf{9},
  36--39\relax
\mciteBstWouldAddEndPuncttrue
\mciteSetBstMidEndSepPunct{\mcitedefaultmidpunct}
{\mcitedefaultendpunct}{\mcitedefaultseppunct}\relax
\EndOfBibitem
\bibitem[Goodwin(2010)]{Goodwin_2010}
A.~L. Goodwin, \emph{Nature Mater.}, 2010, \textbf{9}, 7--8\relax
\mciteBstWouldAddEndPuncttrue
\mciteSetBstMidEndSepPunct{\mcitedefaultmidpunct}
{\mcitedefaultendpunct}{\mcitedefaultseppunct}\relax
\EndOfBibitem
\bibitem[Shepherd \emph{et~al.}(2012)Shepherd, Palamarciuc, Rosa, Guionneau,
  Moln{\'a}r, L{\'e}tard, and Bousseksou]{Shepherd_2012}
H.~J. Shepherd, T.~Palamarciuc, P.~Rosa, P.~Guionneau, G.~Moln{\'a}r, J.-F.
  L{\'e}tard and A.~Bousseksou, \emph{Angew. Chem. Int. Ed.}, 2012,
  \textbf{51}, 3910--3914\relax
\mciteBstWouldAddEndPuncttrue
\mciteSetBstMidEndSepPunct{\mcitedefaultmidpunct}
{\mcitedefaultendpunct}{\mcitedefaultseppunct}\relax
\EndOfBibitem
\bibitem[Duyker \emph{et~al.}(2016)Duyker, Peterson, Kearley, Studer, and
  Kepert]{Duyker_2016}
S.~G. Duyker, V.~K. Peterson, G.~J. Kearley, A.~J. Studer and C.~J. Kepert,
  \emph{Nature Chem.}, 2016, \textbf{8}, 270--275\relax
\mciteBstWouldAddEndPuncttrue
\mciteSetBstMidEndSepPunct{\mcitedefaultmidpunct}
{\mcitedefaultendpunct}{\mcitedefaultseppunct}\relax
\EndOfBibitem
\bibitem[Li \emph{et~al.}(2014)Li, Thirumurugan, Barton, Lin, Henke, Yeung,
  Wharmby, Bithell, Howard, and Cheetham]{LTB2014}
W.~Li, A.~Thirumurugan, P.~T. Barton, Z.~Lin, S.~Henke, H.~H.-M. Yeung, M.~T.
  Wharmby, E.~G. Bithell, C.~J. Howard and A.~K. Cheetham, \emph{{J}. {A}m.
  {C}hem. {S}oc.}, 2014, \textbf{136}, 7801--7804\relax
\mciteBstWouldAddEndPuncttrue
\mciteSetBstMidEndSepPunct{\mcitedefaultmidpunct}
{\mcitedefaultendpunct}{\mcitedefaultseppunct}\relax
\EndOfBibitem
\bibitem[Jones \emph{et~al.}(2014)Jones, Knight, Marshall, Clews, Darton,
  Pyatt, Coles, and Horton]{JKM2014}
R.~H. Jones, K.~S. Knight, W.~G. Marshall, J.~Clews, R.~J. Darton, D.~Pyatt,
  S.~J. Coles and P.~N. Horton, \emph{CrystEngComm}, 2014, \textbf{16},
  237--243\relax
\mciteBstWouldAddEndPuncttrue
\mciteSetBstMidEndSepPunct{\mcitedefaultmidpunct}
{\mcitedefaultendpunct}{\mcitedefaultseppunct}\relax
\EndOfBibitem
\bibitem[Yot \emph{et~al.}(2012)Yot, Ma, Haines, Yang, Ghoufi, Devic, Serre,
  Dmitriev, F{\'e}rey, Zhong, and Maurin]{Yot_2012}
P.~G. Yot, Q.~Ma, J.~Haines, Q.~Yang, A.~Ghoufi, T.~Devic, C.~Serre,
  V.~Dmitriev, G.~F{\'e}rey, C.~Zhong and G.~Maurin, \emph{Chem. Sci.}, 2012,
  \textbf{3}, 1100--1104\relax
\mciteBstWouldAddEndPuncttrue
\mciteSetBstMidEndSepPunct{\mcitedefaultmidpunct}
{\mcitedefaultendpunct}{\mcitedefaultseppunct}\relax
\EndOfBibitem
\bibitem[Engel \emph{et~al.}(2014)Engel, Smith, Bezuidenhout, and
  Barbour]{ESB2014}
E.~R. Engel, V.~J. Smith, C.~X. Bezuidenhout and L.~J. Barbour, \emph{Chem.
  Commun.}, 2014, \textbf{50}, 4238--4241\relax
\mciteBstWouldAddEndPuncttrue
\mciteSetBstMidEndSepPunct{\mcitedefaultmidpunct}
{\mcitedefaultendpunct}{\mcitedefaultseppunct}\relax
\EndOfBibitem
\bibitem[Salles \emph{et~al.}(2010)Salles, Maurin, Serre, Llewellyn,
  Kn{\"o}fel, Choi, Filinchuk, Oliviero, Vimont, Long, and
  F{\'e}rey]{Salles_2010}
F.~Salles, G.~Maurin, C.~Serre, P.~L. Llewellyn, C.~Kn{\"o}fel, H.~J. Choi,
  Y.~Filinchuk, L.~Oliviero, A.~Vimont, J.~R. Long and G.~F{\'e}rey, \emph{J.
  Am. Chem. Soc.}, 2010, \textbf{132}, 13782--13788\relax
\mciteBstWouldAddEndPuncttrue
\mciteSetBstMidEndSepPunct{\mcitedefaultmidpunct}
{\mcitedefaultendpunct}{\mcitedefaultseppunct}\relax
\EndOfBibitem
\bibitem[Goodwin \emph{et~al.}(2008)Goodwin, Keen, Tucker, Dove, Peters, and
  Evans]{GKT2008a}
A.~L. Goodwin, D.~A. Keen, M.~G. Tucker, M.~T. Dove, L.~Peters and J.~S.~O.
  Evans, \emph{{J}. {A}m. {C}hem. {S}oc.}, 2008, \textbf{130}, 9660--9661\relax
\mciteBstWouldAddEndPuncttrue
\mciteSetBstMidEndSepPunct{\mcitedefaultmidpunct}
{\mcitedefaultendpunct}{\mcitedefaultseppunct}\relax
\EndOfBibitem
\bibitem[Ho and Taylor(1998)]{HT1998}
C.~Y. Ho and R.~E. Taylor, \emph{Thermal Expansion of Solids}, ASM
  International, Ohio, 1998\relax
\mciteBstWouldAddEndPuncttrue
\mciteSetBstMidEndSepPunct{\mcitedefaultmidpunct}
{\mcitedefaultendpunct}{\mcitedefaultseppunct}\relax
\EndOfBibitem
\bibitem[Krishnan(1979)]{Krishnan_1979}
\emph{Thermal Expansion of Crystals}, ed. R.~Krishnan, Pergamon Press, Oxford,
  1979, vol.~22\relax
\mciteBstWouldAddEndPuncttrue
\mciteSetBstMidEndSepPunct{\mcitedefaultmidpunct}
{\mcitedefaultendpunct}{\mcitedefaultseppunct}\relax
\EndOfBibitem
\bibitem[Goodwin \emph{et~al.}(2008)Goodwin, Keen, and Tucker]{GKT2008b}
A.~L. Goodwin, D.~A. Keen and M.~G. Tucker, \emph{Proc. Natl. Acad. Sci.,
  U.S.A.}, 2008, \textbf{105}, 18708--18713\relax
\mciteBstWouldAddEndPuncttrue
\mciteSetBstMidEndSepPunct{\mcitedefaultmidpunct}
{\mcitedefaultendpunct}{\mcitedefaultseppunct}\relax
\EndOfBibitem
\bibitem[Greaves \emph{et~al.}(2011)Greaves, Greer, Lakes, and
  Rouxel]{Greaves_2011}
G.~N. Greaves, A.~L. Greer, R.~S. Lakes and T.~Rouxel, \emph{Nature Mater.},
  2011, \textbf{10}, 823--837\relax
\mciteBstWouldAddEndPuncttrue
\mciteSetBstMidEndSepPunct{\mcitedefaultmidpunct}
{\mcitedefaultendpunct}{\mcitedefaultseppunct}\relax
\EndOfBibitem
\bibitem[Panda \emph{et~al.}(2014)Panda, Run{\v c}evski, Chandra~Sahoo, Belik,
  Nath, Dinnebier, and Naumov]{PRC2014}
M.~K. Panda, T.~Run{\v c}evski, S.~Chandra~Sahoo, A.~A. Belik, N.~K. Nath,
  R.~E. Dinnebier and P.~Naumov, \emph{Nature Commun.}, 2014, \textbf{5},
  4811\relax
\mciteBstWouldAddEndPuncttrue
\mciteSetBstMidEndSepPunct{\mcitedefaultmidpunct}
{\mcitedefaultendpunct}{\mcitedefaultseppunct}\relax
\EndOfBibitem
\bibitem[Panda \emph{et~al.}(2016)Panda, Centore, Caus{\`a}, Tuzi, Borbone, and
  Naumov]{PCC2016}
M.~K. Panda, R.~Centore, M.~Caus{\`a}, A.~Tuzi, F.~Borbone and P.~Naumov,
  \emph{Sci. Rep.}, 2016, \textbf{6}, 29610\relax
\mciteBstWouldAddEndPuncttrue
\mciteSetBstMidEndSepPunct{\mcitedefaultmidpunct}
{\mcitedefaultendpunct}{\mcitedefaultseppunct}\relax
\EndOfBibitem
\bibitem[Kor{\v c}ok \emph{et~al.}(2009)Kor{\v c}ok, Katz, and
  Leznoff]{Korcok_2009}
J.~L. Kor{\v c}ok, M.~J. Katz and D.~B. Leznoff, \emph{J. Am. Chem. Soc.},
  2009, \textbf{131}, 4866--4871\relax
\mciteBstWouldAddEndPuncttrue
\mciteSetBstMidEndSepPunct{\mcitedefaultmidpunct}
{\mcitedefaultendpunct}{\mcitedefaultseppunct}\relax
\EndOfBibitem
\bibitem[Ludi and G{\"u}del(1968)]{LG1968}
A.~Ludi and H.~U. G{\"u}del, \emph{Helv. Chim. Acta}, 1968, \textbf{51},
  1762--1765\relax
\mciteBstWouldAddEndPuncttrue
\mciteSetBstMidEndSepPunct{\mcitedefaultmidpunct}
{\mcitedefaultendpunct}{\mcitedefaultseppunct}\relax
\EndOfBibitem
\bibitem[Goodwin \emph{et~al.}(2008)Goodwin, Calleja, Conterio, Dove, Evans,
  Keen, Peters, and Tucker]{GCC2008}
A.~L. Goodwin, M.~Calleja, M.~J. Conterio, M.~T. Dove, J.~S.~O. Evans, D.~A.
  Keen, L.~Peters and M.~G. Tucker, \emph{Science}, 2008, \textbf{319},
  794--797\relax
\mciteBstWouldAddEndPuncttrue
\mciteSetBstMidEndSepPunct{\mcitedefaultmidpunct}
{\mcitedefaultendpunct}{\mcitedefaultseppunct}\relax
\EndOfBibitem
\bibitem[Conterio \emph{et~al.}(2008)Conterio, Goodwin, Tucker, Keen, Dove,
  Peters, and Evans]{CGT2008}
M.~J. Conterio, A.~L. Goodwin, M.~G. Tucker, D.~A. Keen, M.~T. Dove, L.~Peters
  and J.~S.~O. Evans, \emph{J. Phys.: Cond. Matt.}, 2008, \textbf{20},
  255225\relax
\mciteBstWouldAddEndPuncttrue
\mciteSetBstMidEndSepPunct{\mcitedefaultmidpunct}
{\mcitedefaultendpunct}{\mcitedefaultseppunct}\relax
\EndOfBibitem
\bibitem[Baughman \emph{et~al.}(1998)Baughman, Stafstr{\"o}m, Cui, and
  Dantas]{BRS1998}
R.~H. Baughman, S.~Stafstr{\"o}m, C.~Cui and S.~O. Dantas, \emph{Science},
  1998, \textbf{279}, 1522--1524\relax
\mciteBstWouldAddEndPuncttrue
\mciteSetBstMidEndSepPunct{\mcitedefaultmidpunct}
{\mcitedefaultendpunct}{\mcitedefaultseppunct}\relax
\EndOfBibitem
\bibitem[Lakes and Wojciechowski(2008)]{Lakes_2008}
R.~Lakes and K.~W. Wojciechowski, \emph{Phys. Stat. Sol. B}, 2008,
  \textbf{245}, 545--551\relax
\mciteBstWouldAddEndPuncttrue
\mciteSetBstMidEndSepPunct{\mcitedefaultmidpunct}
{\mcitedefaultendpunct}{\mcitedefaultseppunct}\relax
\EndOfBibitem
\bibitem[Cairns and Goodwin(2015)]{Cairns_2015}
A.~B. Cairns and A.~L. Goodwin, \emph{Phys. Chem. Chem. Phys.}, 2015,
  \textbf{17}, 20449--20465\relax
\mciteBstWouldAddEndPuncttrue
\mciteSetBstMidEndSepPunct{\mcitedefaultmidpunct}
{\mcitedefaultendpunct}{\mcitedefaultseppunct}\relax
\EndOfBibitem
\bibitem[Jansen(1987)]{Jansen_1987}
M.~Jansen, \emph{Angew. Chem. Int. Ed. Engl.}, 1987, \textbf{26},
  1098--1110\relax
\mciteBstWouldAddEndPuncttrue
\mciteSetBstMidEndSepPunct{\mcitedefaultmidpunct}
{\mcitedefaultendpunct}{\mcitedefaultseppunct}\relax
\EndOfBibitem
\bibitem[O{'}Grady and Kaltsoyannis(2004)]{OK2004}
E.~O{'}Grady and N.~Kaltsoyannis, \emph{Phys. Chem. Chem. Phys.}, 2004,
  \textbf{6}, 680--687\relax
\mciteBstWouldAddEndPuncttrue
\mciteSetBstMidEndSepPunct{\mcitedefaultmidpunct}
{\mcitedefaultendpunct}{\mcitedefaultseppunct}\relax
\EndOfBibitem
\bibitem[Bessler \emph{et~al.}(2001)Bessler, Calzavara, Deflon, and
  Niquet]{BCD2001}
K.~E. Bessler, L.~A. d.~P. Calzavara, V.~M. Deflon and E.~Niquet, \emph{Acta
  Cryst. E}, 2001, \textbf{57}, m522--m523\relax
\mciteBstWouldAddEndPuncttrue
\mciteSetBstMidEndSepPunct{\mcitedefaultmidpunct}
{\mcitedefaultendpunct}{\mcitedefaultseppunct}\relax
\EndOfBibitem
\bibitem[Hunt \emph{et~al.}(2016)Hunt, Cliffe, Hill, Cairns, Funnell, and
  Goodwin]{Hunt_2015}
S.~J. Hunt, M.~J. Cliffe, J.~A. Hill, A.~B. Cairns, N.~P. Funnell and A.~L.
  Goodwin, \emph{CrystEngComm}, 2016, \textbf{17}, 361--369\relax
\mciteBstWouldAddEndPuncttrue
\mciteSetBstMidEndSepPunct{\mcitedefaultmidpunct}
{\mcitedefaultendpunct}{\mcitedefaultseppunct}\relax
\EndOfBibitem
\bibitem[Pauling and Pauling(1968)]{Pauling_1968}
L.~Pauling and P.~Pauling, \emph{Proc. Natl. Acad. Sci.}, 1968, \textbf{60},
  362--367\relax
\mciteBstWouldAddEndPuncttrue
\mciteSetBstMidEndSepPunct{\mcitedefaultmidpunct}
{\mcitedefaultendpunct}{\mcitedefaultseppunct}\relax
\EndOfBibitem
\bibitem[Haser \emph{et~al.}(1977)Haser, Bonnet, and Roziere]{Haser_1977}
R.~Haser, B.~Bonnet and J.~Roziere, \emph{J. Mol. Structure}, 1977,
  \textbf{40}, 177--189\relax
\mciteBstWouldAddEndPuncttrue
\mciteSetBstMidEndSepPunct{\mcitedefaultmidpunct}
{\mcitedefaultendpunct}{\mcitedefaultseppunct}\relax
\EndOfBibitem
\bibitem[Keen \emph{et~al.}(2010)Keen, Dove, Evans, Goodwin, Peters, and
  Tucker]{KDE2010}
D.~A. Keen, M.~T. Dove, J.~S.~O. Evans, A.~L. Goodwin, L.~Peters and M.~G.
  Tucker, \emph{J. Phys.: Cond. Matt.}, 2010, \textbf{22}, 404202\relax
\mciteBstWouldAddEndPuncttrue
\mciteSetBstMidEndSepPunct{\mcitedefaultmidpunct}
{\mcitedefaultendpunct}{\mcitedefaultseppunct}\relax
\EndOfBibitem
\bibitem[Lind(2012)]{Lind_2012}
C.~Lind, \emph{Materials}, 2012, \textbf{5}, 1125--1154\relax
\mciteBstWouldAddEndPuncttrue
\mciteSetBstMidEndSepPunct{\mcitedefaultmidpunct}
{\mcitedefaultendpunct}{\mcitedefaultseppunct}\relax
\EndOfBibitem
\bibitem[Coelho()]{topas}
A.~A. Coelho, \emph{{TOPAS}-Academic, version 4.1 (computer software)}, Coelho
  software technical report\relax
\mciteBstWouldAddEndPuncttrue
\mciteSetBstMidEndSepPunct{\mcitedefaultmidpunct}
{\mcitedefaultendpunct}{\mcitedefaultseppunct}\relax
\EndOfBibitem
\bibitem[Stephens(1999)]{SP1999}
P.~W. Stephens, \emph{J. Appl. Cryst.}, 1999, \textbf{32}, 281--289\relax
\mciteBstWouldAddEndPuncttrue
\mciteSetBstMidEndSepPunct{\mcitedefaultmidpunct}
{\mcitedefaultendpunct}{\mcitedefaultseppunct}\relax
\EndOfBibitem
\bibitem[Ratuszna and Ma{\l }ecki(2000)]{RM2000}
A.~Ratuszna and G.~Ma{\l }ecki, \emph{Mater. Sci. Forum}, 2000,
  \textbf{321-324}, 947--953\relax
\mciteBstWouldAddEndPuncttrue
\mciteSetBstMidEndSepPunct{\mcitedefaultmidpunct}
{\mcitedefaultendpunct}{\mcitedefaultseppunct}\relax
\EndOfBibitem
\bibitem[Cliffe and Goodwin(2012)]{pascal}
M.~J. Cliffe and A.~L. Goodwin, \emph{J. Appl. Cryst.}, 2012, \textbf{45},
  1321--1329\relax
\mciteBstWouldAddEndPuncttrue
\mciteSetBstMidEndSepPunct{\mcitedefaultmidpunct}
{\mcitedefaultendpunct}{\mcitedefaultseppunct}\relax
\EndOfBibitem
\bibitem[Nye(1957)]{Nye_1957}
J.~F. Nye, \emph{Physical Properties of Crystals}, Oxford University Press,
  Oxford, 1957\relax
\mciteBstWouldAddEndPuncttrue
\mciteSetBstMidEndSepPunct{\mcitedefaultmidpunct}
{\mcitedefaultendpunct}{\mcitedefaultseppunct}\relax
\EndOfBibitem
\bibitem[Blum \emph{et~al.}(2009)Blum, Gehrke, Hanke, Havu, Havu, Ren, Reuter,
  and Scheffler]{Blum_2009}
V.~Blum, R.~Gehrke, F.~Hanke, P.~Havu, V.~Havu, X.~Ren, K.~Reuter and
  M.~Scheffler, \emph{Comput. Phys. Commun.}, 2009, \textbf{180},
  2175--2196\relax
\mciteBstWouldAddEndPuncttrue
\mciteSetBstMidEndSepPunct{\mcitedefaultmidpunct}
{\mcitedefaultendpunct}{\mcitedefaultseppunct}\relax
\EndOfBibitem
\bibitem[Perdew \emph{et~al.}(1996)Perdew, Burke, and Ernzerhof]{Perdew_1996}
J.~P. Perdew, K.~Burke and M.~Ernzerhof, \emph{Phys. Rev. Lett.}, 1996,
  \textbf{77}, 3865--3868\relax
\mciteBstWouldAddEndPuncttrue
\mciteSetBstMidEndSepPunct{\mcitedefaultmidpunct}
{\mcitedefaultendpunct}{\mcitedefaultseppunct}\relax
\EndOfBibitem
\bibitem[Tkatchenko and Scheffler(2009)]{Tkatchenko_2009}
A.~Tkatchenko and M.~Scheffler, \emph{Phys. Rev. Lett.}, 2009, \textbf{102},
  073005\relax
\mciteBstWouldAddEndPuncttrue
\mciteSetBstMidEndSepPunct{\mcitedefaultmidpunct}
{\mcitedefaultendpunct}{\mcitedefaultseppunct}\relax
\EndOfBibitem
\bibitem[Tkatchenko \emph{et~al.}(2012)Tkatchenko, DiStasio~Jr., Car, and
  Scheffler]{Tkatchenko_2012}
A.~Tkatchenko, R.~A. DiStasio~Jr., R.~Car and M.~Scheffler, \emph{Phys. Rev.
  Lett.}, 2012, \textbf{108}, 236402\relax
\mciteBstWouldAddEndPuncttrue
\mciteSetBstMidEndSepPunct{\mcitedefaultmidpunct}
{\mcitedefaultendpunct}{\mcitedefaultseppunct}\relax
\EndOfBibitem
\bibitem[Ambrosetti \emph{et~al.}(2014)Ambrosetti, Reilly, DiStasio~Jr., and
  Tkatchenko]{Ambrosetti_2014}
A.~Ambrosetti, A.~M. Reilly, R.~A. DiStasio~Jr. and A.~Tkatchenko, \emph{J.
  Chem. Phys.}, 2014, \textbf{140}, 18A508\relax
\mciteBstWouldAddEndPuncttrue
\mciteSetBstMidEndSepPunct{\mcitedefaultmidpunct}
{\mcitedefaultendpunct}{\mcitedefaultseppunct}\relax
\EndOfBibitem
\bibitem[Gale(1997)]{gulp}
J.~D. Gale, \emph{J. Chem. Soc.{,} Faraday Trans.}, 1997, \textbf{93},
  629--637\relax
\mciteBstWouldAddEndPuncttrue
\mciteSetBstMidEndSepPunct{\mcitedefaultmidpunct}
{\mcitedefaultendpunct}{\mcitedefaultseppunct}\relax
\EndOfBibitem
\bibitem[Sharpe(1976)]{Sharpe_1976}
A.~G. Sharpe, \emph{The Chemistry of Cyano Complexes of the Transition Metals},
  Academic Press, London, 1976\relax
\mciteBstWouldAddEndPuncttrue
\mciteSetBstMidEndSepPunct{\mcitedefaultmidpunct}
{\mcitedefaultendpunct}{\mcitedefaultseppunct}\relax
\EndOfBibitem
\bibitem[Widmann \emph{et~al.}(2005)Widmann, Kahlert, Wulff, and
  Scholz]{WKW2005}
A.~Widmann, H.~Kahlert, H.~Wulff and F.~Scholz, \emph{J. Sol. St.
  Electrochem.}, 2005, \textbf{9}, 380--389\relax
\mciteBstWouldAddEndPuncttrue
\mciteSetBstMidEndSepPunct{\mcitedefaultmidpunct}
{\mcitedefaultendpunct}{\mcitedefaultseppunct}\relax
\EndOfBibitem
\bibitem[Hibble \emph{et~al.}(2004)Hibble, Eversfield, Cowley, and
  Chippindale]{HEC2004}
S.~J. Hibble, S.~G. Eversfield, A.~R. Cowley and A.~M. Chippindale,
  \emph{Angew. Chem. Int. Ed.}, 2004, \textbf{43}, 628--630\relax
\mciteBstWouldAddEndPuncttrue
\mciteSetBstMidEndSepPunct{\mcitedefaultmidpunct}
{\mcitedefaultendpunct}{\mcitedefaultseppunct}\relax
\EndOfBibitem
\bibitem[Katz \emph{et~al.}(2008)Katz, Sakai, and Leznoff]{KSL2008}
M.~J. Katz, K.~Sakai and D.~B. Leznoff, \emph{Chem. Soc. Rev.}, 2008,
  \textbf{37}, 1884--1895\relax
\mciteBstWouldAddEndPuncttrue
\mciteSetBstMidEndSepPunct{\mcitedefaultmidpunct}
{\mcitedefaultendpunct}{\mcitedefaultseppunct}\relax
\EndOfBibitem
\bibitem[Batsanov(2001)]{B2001}
S.~S. Batsanov, \emph{Inorg. Mater.}, 2001, \textbf{37}, 871--885\relax
\mciteBstWouldAddEndPuncttrue
\mciteSetBstMidEndSepPunct{\mcitedefaultmidpunct}
{\mcitedefaultendpunct}{\mcitedefaultseppunct}\relax
\EndOfBibitem
\bibitem[Hibble \emph{et~al.}(2010)Hibble, Wood, Bibl{\'e}, Pohl, Tucker,
  Hannon, and Chippindale]{HWB2010}
S.~J. Hibble, G.~B. Wood, E.~J. Bibl{\'e}, A.~H. Pohl, M.~G. Tucker, A.~C.
  Hannon and A.~M. Chippindale, \emph{Z. Krist.}, 2010, \textbf{225},
  457--462\relax
\mciteBstWouldAddEndPuncttrue
\mciteSetBstMidEndSepPunct{\mcitedefaultmidpunct}
{\mcitedefaultendpunct}{\mcitedefaultseppunct}\relax
\EndOfBibitem
\bibitem[Mary \emph{et~al.}(1996)Mary, Evans, Vogt, and Sleight]{Mary_1996}
T.~A. Mary, J.~S.~O. Evans, T.~Vogt and A.~W. Sleight, \emph{Science}, 1996,
  \textbf{272}, 90--92\relax
\mciteBstWouldAddEndPuncttrue
\mciteSetBstMidEndSepPunct{\mcitedefaultmidpunct}
{\mcitedefaultendpunct}{\mcitedefaultseppunct}\relax
\EndOfBibitem
\bibitem[Barerra \emph{et~al.}(2005)Barerra, Bruno, Barron, and
  Allan]{Barrera_2005}
G.~D. Barerra, J.~A.~O. Bruno, T.~H.~K. Barron and N.~L. Allan, \emph{J. Phys.:
  Cond. Matt.}, 2005, \textbf{17}, R217--R252\relax
\mciteBstWouldAddEndPuncttrue
\mciteSetBstMidEndSepPunct{\mcitedefaultmidpunct}
{\mcitedefaultendpunct}{\mcitedefaultseppunct}\relax
\EndOfBibitem
\bibitem[Adamson \emph{et~al.}(2015)Adamson, Lucas, Cairns, Funnell, Tucker,
  Kleppe, Hriljac, and Goodwin]{Adamson_2015}
J.~Adamson, T.~C. Lucas, A.~B. Cairns, N.~P. Funnell, M.~G. Tucker, A.~K.
  Kleppe, J.~A. Hriljac and A.~L. Goodwin, \emph{Physica B}, 2015,
  \textbf{479}, 35--40\relax
\mciteBstWouldAddEndPuncttrue
\mciteSetBstMidEndSepPunct{\mcitedefaultmidpunct}
{\mcitedefaultendpunct}{\mcitedefaultseppunct}\relax
\EndOfBibitem
\bibitem[Liu \emph{et~al.}(2016)Liu, Hermann, and Tkatchenko]{Liu_2016}
X.~Liu, J.~Hermann and A.~Tkatchenko, \emph{J. Chem. Phys.}, 2016,
  \textbf{145}, 241101\relax
\mciteBstWouldAddEndPuncttrue
\mciteSetBstMidEndSepPunct{\mcitedefaultmidpunct}
{\mcitedefaultendpunct}{\mcitedefaultseppunct}\relax
\EndOfBibitem
\bibitem[Calleja \emph{et~al.}(2008)Calleja, Goodwin, and Dove]{CGD2008}
M.~Calleja, A.~L. Goodwin and M.~T. Dove, \emph{J. Phys.: Cond. Matt.}, 2008,
  \textbf{20}, 255226\relax
\mciteBstWouldAddEndPuncttrue
\mciteSetBstMidEndSepPunct{\mcitedefaultmidpunct}
{\mcitedefaultendpunct}{\mcitedefaultseppunct}\relax
\EndOfBibitem
\bibitem[Magnko \emph{et~al.}(2002)Magnko, Schweizer, Rauhut, Schutz, Stoll,
  and Werner]{MSR2002}
L.~Magnko, M.~Schweizer, G.~Rauhut, M.~Schutz, H.~Stoll and H.-J. Werner,
  \emph{Phys. Chem. Chem. Phys.}, 2002, \textbf{4}, 1006--1013\relax
\mciteBstWouldAddEndPuncttrue
\mciteSetBstMidEndSepPunct{\mcitedefaultmidpunct}
{\mcitedefaultendpunct}{\mcitedefaultseppunct}\relax
\EndOfBibitem
\bibitem[Fang \emph{et~al.}(2014)Fang, Dove, and Refson]{Fang_2014}
H.~Fang, M.~T. Dove and K.~Refson, \emph{Phys. Rev. B}, 2014, \textbf{90},
  054302\relax
\mciteBstWouldAddEndPuncttrue
\mciteSetBstMidEndSepPunct{\mcitedefaultmidpunct}
{\mcitedefaultendpunct}{\mcitedefaultseppunct}\relax
\EndOfBibitem
\bibitem[Munn(1972)]{Munn_1972}
R.~W. Munn, \emph{J. Phys. C.: Solid State Phys.}, 1972, \textbf{5},
  535--542\relax
\mciteBstWouldAddEndPuncttrue
\mciteSetBstMidEndSepPunct{\mcitedefaultmidpunct}
{\mcitedefaultendpunct}{\mcitedefaultseppunct}\relax
\EndOfBibitem
\bibitem[Dove and Fang(2016)]{Dove_2016}
M.~T. Dove and H.~Fang, \emph{Rep. Prog. Phys.}, 2016, \textbf{79},
  066503\relax
\mciteBstWouldAddEndPuncttrue
\mciteSetBstMidEndSepPunct{\mcitedefaultmidpunct}
{\mcitedefaultendpunct}{\mcitedefaultseppunct}\relax
\EndOfBibitem
\bibitem[Hermet \emph{et~al.}(2013)Hermet, Catafesta, Bantignies, Levelut,
  Maurin, Cairns, Goodwin, and Haines]{Hermet_2013}
P.~Hermet, J.~Catafesta, J.-L. Bantignies, C.~Levelut, D.~Maurin, A.~B. Cairns,
  A.~L. Goodwin and J.~Haines, \emph{J. Phys. Chem. C}, 2013, \textbf{117},
  12848--12857\relax
\mciteBstWouldAddEndPuncttrue
\mciteSetBstMidEndSepPunct{\mcitedefaultmidpunct}
{\mcitedefaultendpunct}{\mcitedefaultseppunct}\relax
\EndOfBibitem
\bibitem[Hill \emph{et~al.}(2015)Hill, Cairns, Lim, Cassidy, Clarke, and
  Goodwin]{Hill_2015}
J.~A. Hill, A.~B. Cairns, J.~J.~K. Lim, S.~J. Cassidy, S.~J. Clarke and A.~L.
  Goodwin, \emph{CrystEngComm}, 2015, \textbf{17}, 2925--2928\relax
\mciteBstWouldAddEndPuncttrue
\mciteSetBstMidEndSepPunct{\mcitedefaultmidpunct}
{\mcitedefaultendpunct}{\mcitedefaultseppunct}\relax
\EndOfBibitem
\bibitem[Cairns \emph{et~al.}(2013)Cairns, Catafesta, Levelut, Rouquette,
  van~der Lee, Peters, Thompson, Dmitriev, Haines, and Goodwin]{Cairns_2013}
A.~B. Cairns, J.~Catafesta, C.~Levelut, J.~Rouquette, A.~van~der Lee,
  L.~Peters, A.~L. Thompson, V.~Dmitriev, J.~Haines and A.~L. Goodwin,
  \emph{Nature Mater.}, 2013, \textbf{12}, 212--216\relax
\mciteBstWouldAddEndPuncttrue
\mciteSetBstMidEndSepPunct{\mcitedefaultmidpunct}
{\mcitedefaultendpunct}{\mcitedefaultseppunct}\relax
\EndOfBibitem
\bibitem[Woodall \emph{et~al.}(2013)Woodall, Beavers, Christensen, Hatcher,
  Intissar, Parlett, Teat, Reber, and Raithby]{Woodall_2013}
C.~H. Woodall, C.~M. Beavers, J.~Christensen, L.~E. Hatcher, M.~Intissar,
  A.~Parlett, S.~J. Teat, C.~Reber and P.~R. Raithby, \emph{Angew. Chem. Int.
  Ed.}, 2013, \textbf{52}, 9691--9694\relax
\mciteBstWouldAddEndPuncttrue
\mciteSetBstMidEndSepPunct{\mcitedefaultmidpunct}
{\mcitedefaultendpunct}{\mcitedefaultseppunct}\relax
\EndOfBibitem
\end{mcitethebibliography}
\bibliographystyle{rsc} 

\end{document}